\documentclass{emulateapj} 

\usepackage[german,english,american]{babel}
\usepackage{enumitem} \usepackage{graphicx} \usepackage{natbib}
\usepackage{color} \usepackage{ifthen}

\definecolor{grey}{rgb}{0.7,0.7,0.7}

\def\lesssim{\mathrel{\hbox{\rlap{\hbox{\lower4pt\hbox{$\sim$}}}\hbox{$<$}}}}
\def\gtrsim{\mathrel{\hbox{\rlap{\hbox{\lower4pt\hbox{$\sim$}}}\hbox{$>$}}}}
\def\propsim{\mathrel{\hbox{\rlap{\hbox{\lower4pt\hbox{$\sim$}}}\hbox{$\propto$}}}}
\newcommand{\vcyg}{V404 Cyg}

\newlength{\figwidth}
\begin{document}

\title{A Joint {\em Chandra} and {\em Swift} view of the 2015
  X-ray Dust Scattering Echo of V404 Cygni}

\author{S. Heinz\altaffilmark{1}}
\email{heinzs@astro.wisc.edu}
\author{L. Corrales\altaffilmark{2}}
\author{R. Smith\altaffilmark{3}}
\author{W.N. Brandt\altaffilmark{4,5,6}}
\author{P.G. Jonker\altaffilmark{7,8}}
\author{R.M. Plotkin\altaffilmark{9,10}}
\author{J. Neilsen\altaffilmark{2,11}}

\altaffiltext{1}{Department of Astronomy, University of Wisconsin-Madison,
  Madison, WI 53706, USA} 
\altaffiltext{2}{Kavli Institute for Astrophysics and Space Research,
  Massachusetts Institute of Technology, Cambridge, MA 02139, USA}
\altaffiltext{3}{Harvard--Smithsonian Center for Astrophysics,
  Cambridge, MA~02138, USA}
\altaffiltext{4}{Department of Astronomy \& Astrophysics, The
  Pennsylvania State University, University Park, PA 16802, USA}
\altaffiltext{5}{Institute for Gravitation and the Cosmos, The Pennsylvania
 State University, University Park, PA 16802, USA}
\altaffiltext{6}{Department of Physics, The Pennsylvania State University,
 University Park, PA 16802, USA}
\altaffiltext{7}{SRON, Netherlands Institute for Space Research, 3584
  CA, Utrecht, the Netherlands}
\altaffiltext{8}{Department of Astrophysics/IMAPP, Radboud University
  Nijmegen, 6500 GL, Nijmegen, The Netherlands}
\altaffiltext{9}{International Centre for Radio Astronomy Research
  (ICRAR), Curtin University, G.P.O. Box U1987, Perth, WA 6845,
  Australia}
\altaffiltext{10}{Department of Astronomy, University of Michigan, 1085 South University Ave, Ann Arbor, MI 48109, USA}
\altaffiltext{11}{Hubble Fellow}

\begin{abstract}
  We present a combined analysis of the {\em Chandra} and {\em Swift}
  observations of the 2015 X-ray echo of V404 Cygni. Using stacking
  analysis, we identify eight separate rings in the echo. We
  reconstruct the soft X-ray lightcurve of the June 2015 outburst
  using the high-resolution {\em Chandra} images and
  cross-correlations of the radial intensity profiles, indicating that
  about 70\% of the outburst fluence occurred during the bright flare
  at the end of the outburst on MJD 57199.8.  By deconvolving the
  intensity profiles with the reconstructed outburst lightcurve, we
  show that the rings correspond to eight separate dust concentrations
  with precise distance determinations. We further show that the
  column density of the clouds varies significantly across the field
  of view, with the centroid of most of the clouds shifted toward the
  Galactic plane, relative to the position of V404 Cyg, invalidating
  the assumption of uniform cloud column typically made in attempts to
  constrain dust properties from light echoes.  We present a new {\tt
    XSPEC} spectral dust scattering model that calculates the
  differential dust scattering cross section for a range of commonly
  used dust distributions and compositions and use it to jointly fit
  the entire set of {\em Swift} echo data. We find that a standard
  Mathis-Rumpl-Nordsieck model provides an adequate fit to the
  ensemble of echo data. The fit is improved by allowing steeper dust
  distributions, and models with simple silicate and graphite grains
  are preferred over models with more complex composition.
\end{abstract}


\section{Introduction}
\label{sec:introduction}
X-ray transients in outburst are among the brightest X-ray objects in
the sky. When such an outburst has a sharp temporal decline and the
transient is located in the plane of the Galaxy, behind a significant
column of dust and gas, X-ray scattering by intervening interstellar
dust grains can generate a bright light echo in the form of rings that
grow in radius with time since the end of the outburst.

Three bright echoes from Galactic X-ray sources have been found to
date: in 2009 from the magnetar 1E 1547.0-5408 \citep{tiengo:10}, in
2014 from the young neutron star X-ray binary Circinus X-1
\citep{heinz:15}, and in 2015 from the black hole X-ray binary V404
Cygni \citep{beardmore:15,beardmore:16}.  The soft gamma-ray repeater
SGR 1806-20 \citep{svirski:11} and the fast X-ray transient IGR
J17544-2619 \citep{mao:14} have also been claimed to show resolved
(but weak) ring echoes, and \citet{mccollough:13} found scattering
echoes from a single Bok globule toward Cygnus X-3.

In this paper, we will present an in-depth analysis of the combined
{\em Chandra} and {\em Swift} data of the July-August 2015 light echo
from V404 Cygni.

\subsection{Light Echoes}
Dust scattering of X-rays from bright point sources is a well-known
and broadly studied phenomenon.  For steady sources, scattering leads
to the formation of a diffuse dust scattering halo around the source
on arcminute scales at soft X-ray energies
\citep{mauche:86,mathis:91,predehl:95}. Typically, much of the dust
along the line of sight toward a source in the Galactic plane will be
located in dense molecular clouds, with each cloud contributing to the
total scattering intensity. When the X-ray source exhibits
time-variable behavior, the scattered emission will reflect the
variability of the source, and because the scattered X-rays traverse a
longer path than the X-rays directly received from the source, the
scattered emission will be delayed, creating an echo of the X-ray
variability signatures of the source.  This behavior can be used to
study both the source and the intervening dust
\citep[e.g.][]{xiang:11,corrales:15b}.

\begin{figure}[t]
  \center\resizebox{\columnwidth}{!}{\includegraphics{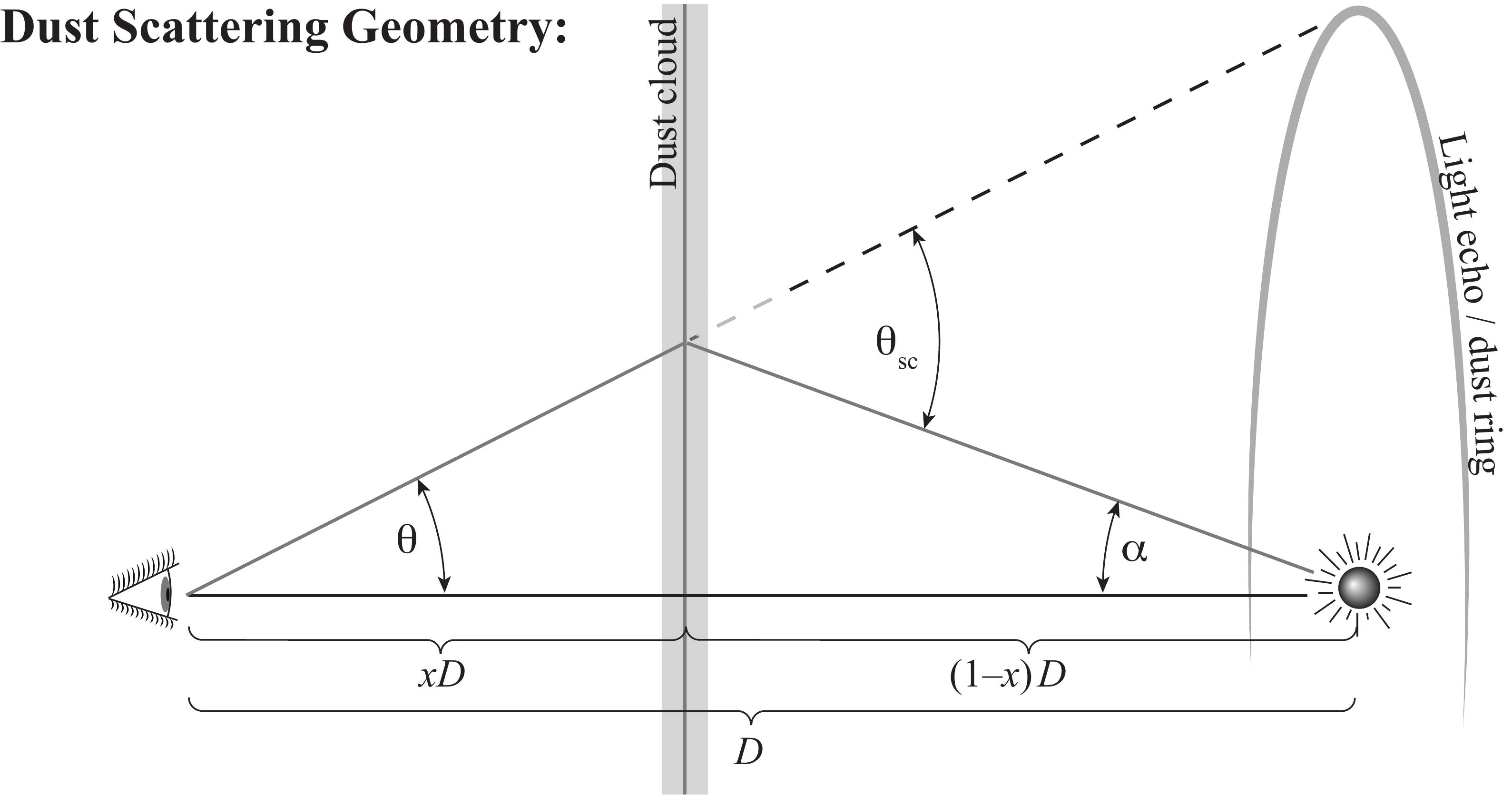}}
  \caption{Cartoon of dust scattering geometry. X-rays from a source
    at distance $D$ are scattered off a dust layer at distance
    $xD$. The observed angle of the scattered X-rays is $\theta$,
    while the true scattering angle is
    $\theta_{\rm sc} =\theta/(1-x)$.}\label{fig:cartoon}\vspace*{6pt}
\end{figure}

If the source exhibits a temporally well-defined flare followed by a
period of quiescence, the scattering signal takes the form of discrete
rings.  The geometry of dust scattering echoes is described in detail
by, e.g., \citet{vianello:07,tiengo:10,heinz:15}. Here, we will
briefly summarize the basic geometric features of light-echo rings and
define quantities used throughout the paper. The cartoon shown in
Fig.~\ref{fig:cartoon} shows a simple sketch of the geometry.

For an X-ray source at distance $D$, X-rays traveling toward the
observer can be scattered by intervening dust. A dust cloud at
distance $D_{\rm dust}=xD$ (where $x$ is the fractional dust distance
relative to the source distance) can scatter X-rays traveling along
some initial angle $\alpha$ relative to the line of sight toward the
observer, such that they arrive at an observed angle $\theta$ relative
to the line of sight. The scattering angle is thus
$\theta_{\rm sc} = \alpha + \theta$.

Because the scattered X-rays (observed at time $t_{\rm obs}$) have to
traverse a longer distance, they will arrive with a time delay
$\Delta t = t_{\rm obs} - t_{\rm flare}$ relative to the un-scattered
X-rays (observed at time $t_{\rm flare}$), given by
\begin{equation}
  \Delta t = \frac{xD\theta^2}{2c\left(1 - x\right)}
  \label{eq:deltat}
\end{equation}
If the dust along the line of sight is concentrated into dense clouds
(such that the cloud extent is small compared to $D$), and if the
X-ray flare is comparable to or shorter than $\Delta t$, the light
echo will take the form of well defined rings with angular distance
from the optical axis toward the source of
\begin{equation}
  \theta = \sqrt{\frac{2c\Delta t\left(1 - x\right)}{xD}}
  \label{eq:theta}
\end{equation}

The (unabsorbed) intensity of the light echo is given by
\citep[e.g.][]{mathis:91}
\begin{equation}
  I_{\nu} = N_{\rm H} \frac{d\sigma_{\rm sc,\nu}}{d\Omega}
  \frac{F_{\nu}\left(t=t_{\rm obs}-\Delta t\right)}{\left(1 -
      x\right)^2}
  \label{eq:intensity}
\end{equation}
where $d\sigma_{\rm sc}/d\Omega$ is the differential dust scattering
cross section (per hydrogen atom), $F_{\nu}(t)$ is the flux of the
flare at time $t$, and $N_{\rm H}$ is the hydrogen column density of
the cloud responsible for the echo.  Photo-electric absorption will
attenuate this intensity by a factor
$\exp{[-\sigma_{\rm ph,\nu}N_{\rm H,tot}]}$, where $N_{\rm H,tot}$ is
the total hydrogen column along the path of the echo and
$\sigma_{\rm ph,\nu}$ is the total photo-electric absorption cross
section at frequency $\nu$.

{\color{black}{For a short flare (i.e., flare duration
$\delta t_{\rm flare} \ll \Delta t$ such that the ring is narrow and
$d\sigma/d\Omega$ can be approximated as constant across the ring),
the flux density from each ring (produced by the entire echo of a
single cloud) at a given energy can be derived by integrating
eq.~(\ref{eq:intensity}) over $\theta$, holding $d\sigma/d\Omega$
constant:
\begin{equation}
  F_{\rm ring,\nu} = \frac{2\pi c N_{\rm H}}{x(1-x)D}\frac{d\sigma_{\rm
      sc,\nu}}{d\Omega} {\mathcal F}_{\nu}
  \label{eq:fluence}
\end{equation}
where ${\mathcal F} \equiv \int dt F$ is the fluence of the flare and
$N_{\rm H}$ is the column density averaged over the entire ring.  Equation
(\ref{eq:fluence}) can be written for ring sections by integrating $I$
over a range in azimuthal angle $\phi \leq 2\pi$ to account for the
fact that $N_{\rm H}$ can vary across the cloud, as discussed further
in \S\ref{sec:spectral_model}.}}

The scattering cross section depends strongly on scattering angle,
with 
\begin{eqnarray}
  \frac{d\sigma(\theta_{\rm sc},E)}{d\Omega} & = & \int_{a_{\rm min}}^{a_{\rm max}}da\,\frac{dN}{da}
    \frac{d\sigma(\theta_{\rm sc},E,a)}{d\Omega} \nonumber \\
  & \sim & \frac{d\sigma}{d\Omega}_{1000,1keV}
    \left(\frac{\theta_{\rm sc}}{1000''}\right)^{-\alpha}
    \left(\frac{E}{1 keV}\right)^{-\beta}
\end{eqnarray}
{\color{black}{with $\alpha \sim 3-4$ and $\beta \sim 3-4$
    \citep[e.g.][]{draine:03}.}}
Since the scattering angle is simply related to $\Delta t$, $x$, and
$\theta$ by
\begin{equation}
  \theta_{\rm sc}=\frac{\theta}{1- x} = \sqrt{\frac{2c\Delta
    t}{xD\left(1-x\right)}}
\label{eq:theta_scatter}
\end{equation}
the intensity and the flux of the light echo decrease roughly as 
\begin{equation}
  F_{\rm ring} \propto \Delta t^{-\alpha/2}
  \label{eq:intensitydecrease}
\end{equation}
with time delay $\Delta t$ between the flare and the time of the
observation.

Observing X-ray light echoes can be used to study properties of the
source as well as the intervening dust. For example, if the distance
to the source is not known, one may use light echoes in combination
with kinematic information from molecular gas to constrain the source
distance \citep{predehl:00,heinz:15}.

On the other hand, if the source distance is known, the echo becomes a
powerful probe of the distribution of interstellar dust, because
eq.~(\ref{eq:theta}) can be solved for the dust distance $x$:
\begin{equation}
  x=\frac{1}{1 + \frac{D\theta^2}{2c\Delta t}}
  \label{eq:x}
\end{equation}
thus allowing 3D maps of the interstellar dust toward the source to be
{\color{black}{constructed.}}

One may also constrain the properties of the X-ray dust scattering
cross section \citep{tiengo:10}. With good temporal coverage of the
light echo, one can determine the dependence of the cross section on
scattering angle (and thus constrain grain size distributions and
scattering physics), and, with an accurate light curve of the outburst
and an independent measure of the hydrogen column density toward the
source, one may determine the absolute value of the scattering cross
section per hydrogen atom.

Because the echo also contains information about the light curve of
the outburst, it may conversely be used to set contraints on the
fluence and temporal evolution of the outburst.

\subsection{The July 2015 V404 Cyg Light Echo}

V404 Cyg is a classical X-ray transient, hosting a
$9^{+0.2}_{-0.6}\,M_{\odot}$ black hole in orbit with a K3 III
companion \citep{khargharia:10}.  The distance of
$D_{\rm V404} = 2.39\pm 0.14\,{\rm kpc}$ to the source is known from
VLBI parallax \citep{miller-jones:09} with high precision.

\begin{figure}[t]
  \center\resizebox{0.9\columnwidth}{!}{\includegraphics{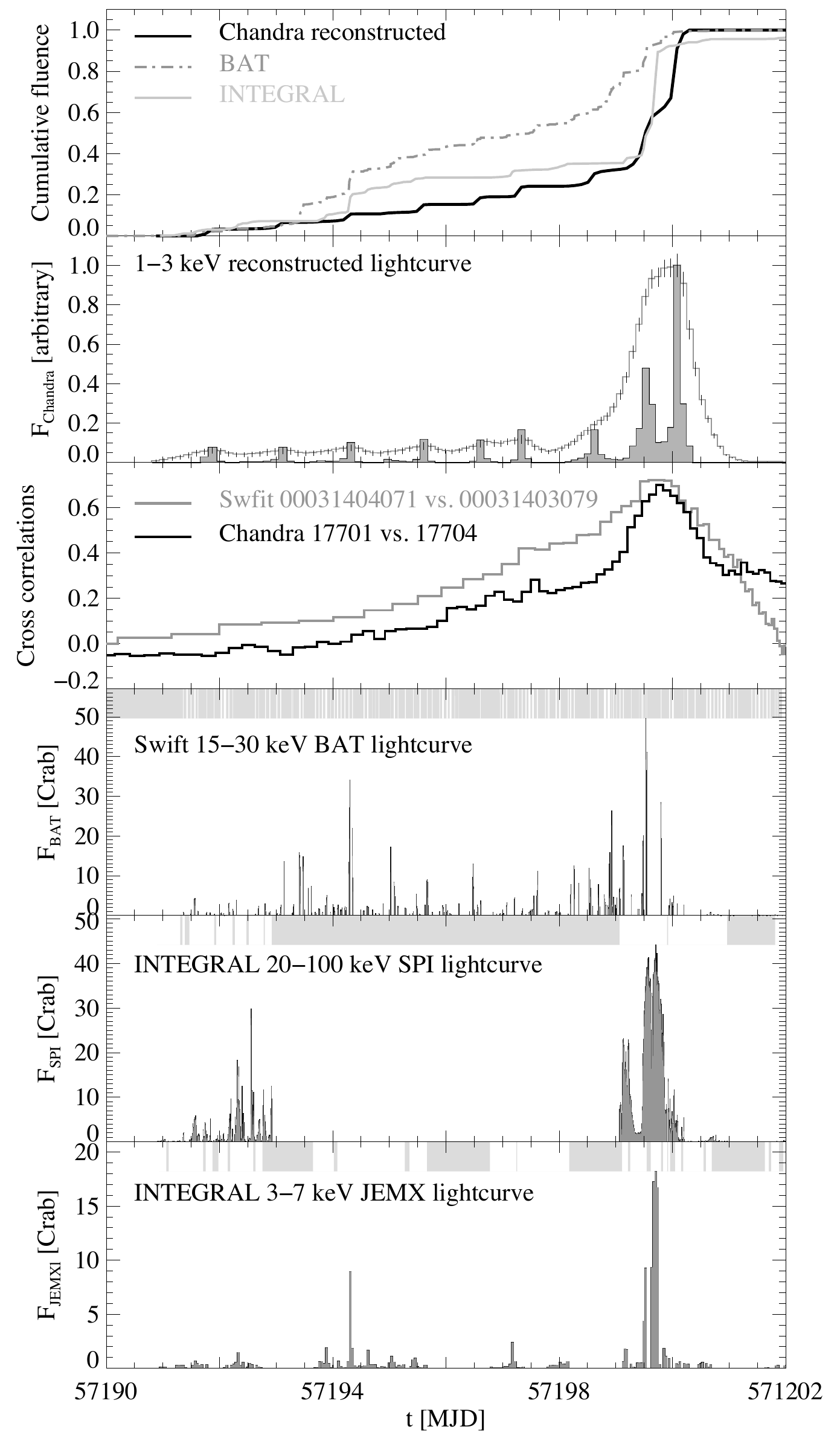}}
  \caption{Lightcurve constraints; from bottom to top:{\em bottom
      panel:} {\em INTEGRAL} JEMX 3-7 keV lightcurve of the June 2015
    outburst in crab units plotted against MJD. Broken gray bar at top
    shows coverage, with gray areas showing gaps in coverage;
    {\color{black}{{\em second panel from bottom:} {\em INTEGRAL} SPI
        20-100 keV lightcurve and coverage fraction (gray bars) of the
        June 2015 outburst in crab units plotted against MJD; {\em
          third panel from bottom:} {\em Swift} BAT 15-30 keV
        lightcurve of the outburst, same convention as bottom panel;
        {\em third panel from top:}}} cross-correlations of {\em
      Chandra} ObsID 17701 vs.~17704 (black) and {\em Swift} ObsID
    00031403071 vs. 00031403079 (gray) plotted against time showing
    the peak of the outburst occurred on MJD 57199.8; {\em second
      panel from top:} Reconstructed lightcurve of the outburst from
    analysis of rings [a] and [b] of Chandra ObsID17704 (light gray
    curve), and the lightcurve deconvolved with the Chandra PSF
    (filled dark gray curve); {\em top panel:} Cumulative fluence of
    the outburst from {\em INTEGRAL} (light gray dash-triple-dot),
    {\em Swift} BAT (dark gray dash-dot) and reconstructed from {\em
      Chandra} (black), indicating that about 70\% of the fluence from
    the outburst was due to the flare on MJD
    57199.8.\label{fig:lightcurve}}
\end{figure}

After 26 years in quiescence, V404 Cyg went into outburst in June 2015
\citep{barthelmy:15,negoro:15,ferrigno:15,rodriguez:15,king:15}, with
peak fluxes in excess of 10 Crab at hard X-ray energies. Shortly after
the end of the outburst, a bright light echo in the form of several
rings was observed in {\em Swift} images of the system.  The echo was
discovered and first reported by \citet{beardmore:15}.  Both {\em
  Swift} and {\em Chandra} followed the temporal evolution of the
light echo.  An analysis of the {\em Swift} data of the echo was
presented by \citet{vasilopoulos:16}, see also \citet{beardmore:16}.

In this paper, we present a combined, in-depth analysis of the {\em
  Chandra} and {\em Swift} data of the echo, introducing a new
spectral modeling code to fit dust scattering echoes using {\tt
  XSPEC}.  Throughout the paper, we will use a default peak time of
the outburst (the time of the brightest peak that contains the largest
portion of the outburst fluence and is responsible for most of the
echo and the bright rings observed) of
\begin{equation}
  t_{\rm flare} = {\rm MJD}\,57199.79
\end{equation}
and delay times of observations $\Delta t$ will be referenced to this
time and to the delay time between $t_{\rm flare}$ and the time of the
first {\em Swift} observation, ObsID 00031403071,
\begin{equation}
  \Delta t_{\rm 0} = 3.68\,{\rm d}
\end{equation}
We will motivate this choice of $t_{\rm flare}$ in
\S\ref{sec:lightcurve}.  Position angles are measured clockwise from
North in FK5 equatorial coordinates.  Unless otherwise specified,
quoted uncertainties are 3-sigma confidence intervals.

The paper is organized as follows: in \S\ref{sec:analysis} we discuss
the reduction and basic analysis of the observations used in the
paper. Section \ref{sec:image_analysis} presents an analysis of the
intensity profiles of the observations.  In \S\ref{sec:spectra}, we
present a new spectral modeling code for dust scattering signals and
describe the spectral fitting technique used to jointly analyze the
entire dust echo. Section \ref{sec:discussion} discusses the results
of the analysis and compares them with previous works. Section
\ref{sec:conclusions} presents conclusions and a summary of our
results.

\section{Data Reduction and Analysis}
\label{sec:analysis}

\subsection{The {\em Swift} BAT and {\em INTEGRAL} JEMX X-ray Light
  Curves of the June 2015 outburst}

A reliable analysis of the V404 Cyg light echo requires accurate
knowledge of the light curve and fluence of the outburst that caused
the echo. Because V404 Cyg was extremely time-variable, with very
short, bright flares \citep{segreto:15,kuulkers:15}, such a light curve
would require full-time monitoring of the object.  No single instrument
operating during the flare observed the source with sufficiently
complete coverage (even if combined) to provide such a light curve.

In particular, {\em MAXI}, the all-sky monitor aboard the
International Space Station and the only dedicated soft (2-4 keV)
X-ray monitor operating at the time, only observed the source with an
uncalibrated instrument (GSC3) and no reliable light curve of the
outburst is available \citep{negoro:15}.

The Burst-Alert-Telescope (BAT) aboard the {\em Swift} spacecraft
provided frequent hard-X-ray observations of the source in June
2015. The BAT light curve of the flare \citep{segreto:15} is plotted
{\color{black}{in the third panel from the bottom}} of
Fig.~\ref{fig:lightcurve}, showing the highly variable behavior of the
source. Plotted in gray above the light curve are the gaps due to
visibility constraints given the low-Earth orbit of the spacecraft,
bad data, and the pointing prioritization of other transient objects,
indicating that the 14\% duty cycle of the BAT was relatively small,
and that many of the very short bright flares were likely missed.  In
consequence, the BAT light curve is insufficient for modeling the
echo.

{\em INTEGRAL}'s JEMX instruments performed dedicated monitoring of
V404 Cyg during June 2015
\citep{rodriguez:15,ferrigno:15,kuulkers:15}. The segment-averaged 3-7
keV light curve is plotted in the bottom panel of
Fig.~\ref{fig:lightcurve}, showing three bright flares as well as
ongoing lower level activity through the period from MJD 57192 to
57200.  The lightcurve was generated using standard pipeline
processing of all {\em INTEGRAL} data taken during the outburst using
the {\tt OSA} software version 10.2.

The coverage by the telescope is continuous for most of the 72 hour
orbital period of the spacecraft, however, the temporal coverage
(again plotted as gray and white bars above the light curve) shows
significant gaps during near-Earth passage, many overlapping with the
gaps in BAT coverage, with a duty cycle of 40\%, giving an approximate
combined duty cycle of 48\% of both telescopes.

{\color{black}{{\em INTEGRAL}'s hard X-ray SPI telescope also observed
    the source during the flare \citep{rodriguez:15,kuulkers:15}; the
    20-100 keV lightcurve is shown in panel 2 of
    Fig.~\ref{fig:lightcurve}, displaying the main flare as well as
    what appears to be a hard pre-cursor not visible in the JEMX
    lightcurve.  The SPI lightcurve does not cover most of the time
    prior to the main flare.}}

Consequently, a full reconstruction of the hard X-ray lightcurve is
not possible from the available coverage.  Furthermore, the analysis
of the echo requires knowledge of the {\em soft} X-ray lightcurve (1-3
keV).  No instrument observed the source continuously in that
band. However, as we will show in \S\ref{sec:lightcurve}, it is
possible to reconstruct the soft X-ray lightcurve of the outburst
(with relatively low temporal resolution) from the {\em Chandra}
images of the inner rings of the echo.

\subsection{Chandra}

{\em Chandra} observed V404 Cyg on 07-11-2015 for 39.5 ksec (ObsID
17701) and on 07-25-2015 for 28.4 ksec (ObsID 17704), as listed in
Table~\ref{tab:obstable}. ObsID 17701 was taken with the High Energy
Transmission Grating (HETG) in place, using ACIS CCDs 5,6,7,8, and 9,
while ObsID 17704 was performed in full-frame mode without gratings,
using ACIS CCDs 2,3,5,6,7, and 8. Because of the potential threat of
lasting chip damage a re-flaring of the point source would have posed
to ACIS\footnote{{\color{black}{For a source at several tens of Crab
      like V404 Cygni at its peak, the accumulated number of counts in
      a given pixel over a 30ksec observation will exceed the
      single-observation dose limit of 625,000 counts set on page 143
      of the Cycle 18 Proposer's Observatory Guide, requiring
      mitigation. Even an increased dither amplitude in such a case
      will not eliminate the potential for damage, leaving placement
      of the source off the chip as the only secure means to avoid
      damage.}}}, the point source in ObsID 17704 was placed in the
chip gap between ACIS S and ACIS I using a SIM Offset of 17.55mm, and
Y- and Z-Offsets of 1'.0 and -0'.3, respectively.

We reduced the data using {\tt CIAO} and {\tt CALDB} version 4.7.
Because of the very small number of point sources visible in the field
of view, we relied on the {\em Chandra} aspect solution of the
observations to align the images. The accuracy of the aspect solution
is typically much better than one arcsecond and sufficient for the
purposes of this analysis.

\begin{figure}[t]
  \center\resizebox{\columnwidth}{!}{\includegraphics{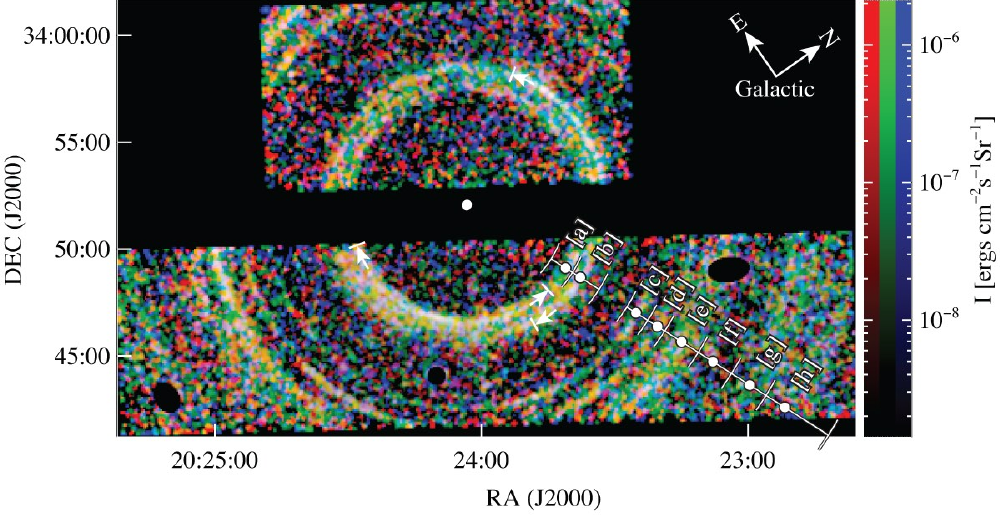}}
  \caption{Three-color exposure-corrected image of {\em Chandra} obsID
    17704 in the 0.5-1,1-2, and 2-3 keV bands (red, green, and blue,
    respectively), smoothed with a 10".5 FWHM Gaussian. Arrow heads
    denote the brightest arclets of the inner two rings [a] and
    [b]. {\color{black}{Overlaid are marks and labels denoting the
        location and spatial extent of spectral extraction regions of
        the eight rings identified in this
        paper.}}}\label{fig:chandra_image}
\end{figure}

ObsID 17704 was background subtracted using the standard blank sky
fields as described in the {\tt CIAO} thread and \citet{hickox:06},
matching hard counts above 10 keV in each CCD separately with the
corresponding blank sky background files to calculate the effective
exposure correction of the background. A three color image in the
bands 0.5-1, 1-2, and 2-3 keV, smoothed with a 10''.5 Gaussian, is
shown in Fig.~\ref{fig:chandra_image}.  The image was extracted in
equatorial (J2000) coordinates and also shows a compass with the axes
of Galactic coordinates for orientation (Galactic North is at a
position angle of about 55$^{\circ}$).

Background point sources in all images were identified using the {\tt
  wavdetect} code in the {\tt CIAO} package \citep{freeman:02} and
removed before further processing.  The image shows six clearly
resolved, distinct rings of dust scattering emission which we label
[a], [b], [c], [d], [e], and [g] from inside out in the analysis below
(see corresponding labels in Fig.~\ref{fig:stacked}).

\begin{figure}[t]
  \center\resizebox{\columnwidth}{!}{\includegraphics{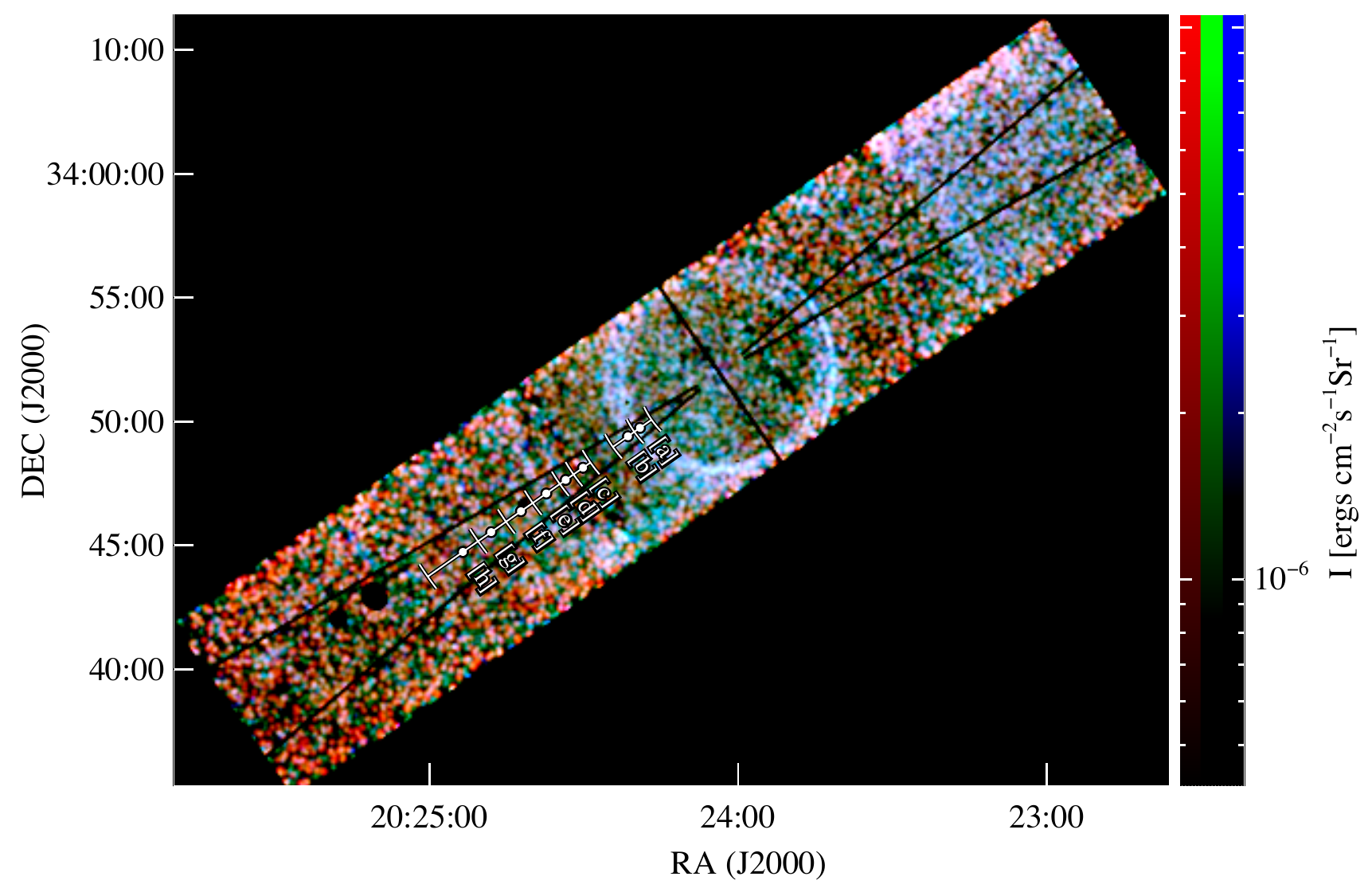}}
  \caption{Three-color exposure-corrected image of {\em Chandra} obsID
    17701 in the 0.5-1,1-2, and 2-3 keV bands (red, green, and blue,
    respectively), smoothed with a 10".5 FWHM
    Gaussian.}\label{fig:17701_image}
\end{figure}

Because no blank background files exist for HETG data, and because the
presence of the HETG significantly affects the background rates, we
were not able to derive accurate backgrounds for ObsID 17701. Neither
stowed nor blank background files produce residual-free
background-subtracted images, which we attribute to the fact that soft
protons focused by the mirrors are affected by the presence of the
gratings, implying that neither blank nor stowed files are appropriate
to use for HETG observations. Figure \ref{fig:17701_image} shows a
three color image of ObsID 17701 without background subtraction.  Only
the innermost rings of the echo, rings [a] and [b] in our notation
below, are sufficiently above the background in ObsID 17701 to discern
them by eye in the image.

The rings in the {\em Chandra} images are very sharp. For further
analysis, we constructed radial intensity profiles for both ObsID
17701 and 17704 following the procedure outlined in \citet{heinz:15},
using 1000 logarithmically spaced radial bins between 3 and 20
arcminutes from the location of V404 Cyg.

The point-source position for ObsID 17701 is coincident with the known
position of V404 Cyg.  As mentioned above, for instrument safety
reasons, the point source was placed in the chip gap between ACIS S
and ACIS I in ObsID 17704.  Using the profiles in five different 15
degree segments in the North-West quadrant of the inner rings [a] and
[b] (all located on ACIS CCD 6), we verified that the centroid of the
rings is coincident with the known position of V404 Cyg to within
about an arcsecond, indicating that the {\em Chandra} aspect solution
is accurate.  No further reprojection of the aspect for ObsID 17704
was therefore needed.

\subsection{Swift}

For this work, we analyzed 50 separate {\em Swift} XRT imaging data
sets of the light echo taken in photon-counting mode, as listed in
Table \ref{tab:obstable}.  We further used 33 pre-flare data sets
(ObsID 00031403002 to ObsID 00031403034), which we merged to construct
a single clean sky background events file for subtraction in spectral
analysis (not listed in Table \ref{tab:obstable}).  The table lists
the fiducial time delay between the mean observation time and
$t_{\rm flare}$.  Data were pipeline processed using the standard {\em
  Swift} package in the {\tt HEASOFT} distribution, version 6.17.

All pre-burst data were taken in 2012 during a previous monitoring
campaign when the source was at a low flux. We stacked all pre-burst
observations for a total exposure time of 138,716 seconds and searched
for point sources. All identified point sources were excluded from
spectral analysis. 

Analysis of the pre- and post-outburst sets of the {\em Swift}
observations indicates that attitude reconstruction of {\em all} event
data incorrectly places the point source at the position
{\color{black}{(RA=20:24:03.970,DEC=+33:51:55.33)}}, compared to the
known source location of
{\color{black}{(RA=20:24:03.83,DEC=+33:52:02.2)}}, offset by 7'',
significantly more than the typical 3'' single-observation pointing
uncertainty. This suggests that the star tracker information used in
attitude reconstruction may contain inaccurate stellar
identifications. All {\em Swift} data were subsequently processed
using a corrected source centroid determined from the
emission-weighted position of the source from all 2012
observations. It is worth noting that the {\em Swift} point-source
catalog \citep{evans:14} contains a source at the off-set position
identified as 1SXPS J202404.2+335155 (derived from the 2012
observations) that is identical with V404 Cyg (identified properly in
other catalogs).

Images and exposure maps were generated using standard {\tt ftools}
reduction tools. Background subtraction of spectra was performed using
the sky background generated from the 2012 data.  

{\color{black}{Given its low-Earth orbit, the {\em Swift} particle background is both
low and approximately constant with time.  We verified that the
diffuse hard (5-10 keV) background levels in both the 2012 and 2015
data are consistent with each other to better than 5\% both in
spectral shape and spatial distribution, within the typical Poisson
uncertainties of the individual 2015 observations.

The soft 0.5-5 keV background (relevant for the analysis of the soft
dust scattering echo) is dominated by astronomical sources, thus,
background removal from the stacked sky file is appropriate even if
the non-astronomical (particle) backgrounds would have changed moderately
between 2012 and 2015.}}

\begin{figure}[t]
  \center\resizebox{\columnwidth}{!}{\includegraphics{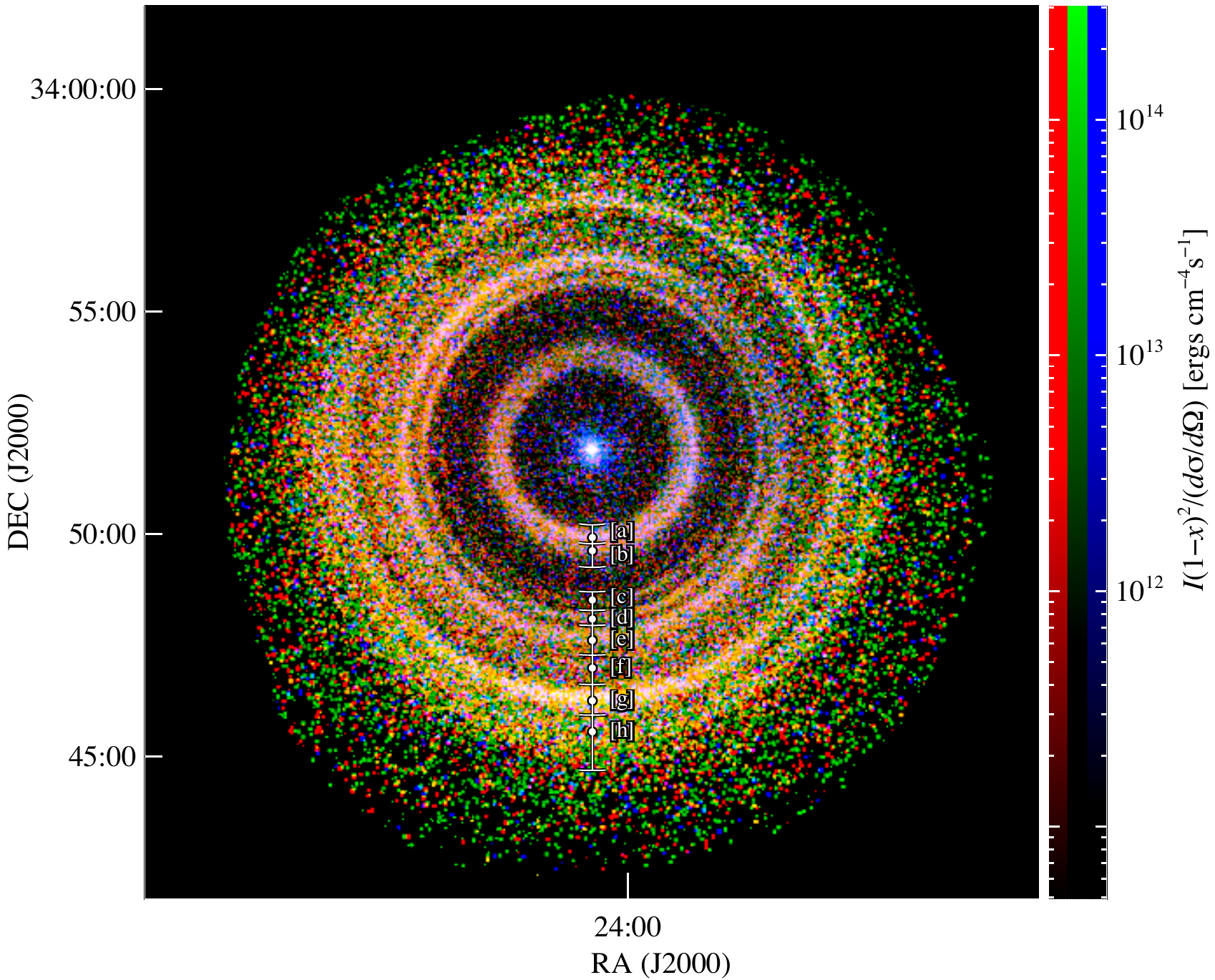}}
  \caption{Three-color stacked exposure corrected {\em Swift} XRT
    image of the 2015 V404 Cyg light echo in the 0.5-1,1-2, and 2-3
    keV bands (red, green, and blue, respectively). All 50 {\em Swift}
    observations listed in Table \ref{tab:obstable} were combined by
    scaling each image to the angular scale of the echo during the
    first observation, ObsID 00031403071 (using eq.~\ref{eq:theta})
    and corrected for temporal attenuation by a factor of
    $(1-x)^2(d\sigma/d\Omega)$ (cross section calculated according to
    the best fit in \S\ref{sec:fit_results}).}\label{fig:stacked}
\end{figure}

To allow a single high-resolution image of the echo to be constructed,
and for comparison with the {\em Chandra} data, we implemented a new
echo stacking procedure.  Figure~\ref{fig:stacked} shows a stacked
color image of all {\em Swift} exposures in the bands
[0.5-1.0,1.0-2.0,2.0-3.0] keV.  This image was constructed as follows:
{\color{black}{Each background-subtracted counts image and each
    exposure map was re-scaled in angular size by a factor
    $\delta_{\theta}\equiv\sqrt{\Delta t_{\rm 0}/\Delta t_{\rm obs}}$,
    centered on the position of V404 Cyg, to match the angular size of
    the echo in {\em Swift} ObsID 00031403071 (using equation
    \ref{eq:theta}), thus matching the peak position of each ring.
    That is, a photon at a given angle $\theta$ is re-mapped to a
    smaller angle $\theta_{0}=\theta\delta_{\theta}$ to match ObsID
    00031403071.}}

In the construction of this image, we implicitly approximated the
lightcurve of the outburst as a single peak (see
\S\ref{sec:lightcurve} for further discussion) to calculate the
fractional dust distance $x$ and scattering angle $\theta_{\rm sc}$
for every pixel of the image, {\color{black}{using eqs.~(\ref{eq:x})
    and (\ref{eq:theta_scatter}), respectively}}.  Each pixel in each
of the counts images was then multiplied by the normalization factor
$(1 - x)^2/[d\sigma(\theta_{\rm sc})/d\Omega]$, using our best-fit
MRN1 dust model to calculate $d\sigma/d\Omega$ (see
\S\ref{sec:crosssection} for details). The image brightness is thus
proportional to $e^{-\tau_{N_{\rm H}}}N_{\rm H}{\mathcal F}$ (i.e., no
correction for photo-electric absorption was made in the
image). {\color{black}{All scaled counts images were then stacked and
    divided by the stacked exposure map.}}  The outer regions of the
image become noise dominated since the total exposure used for those
regions is significantly shorter than the central region.

\section{Image analysis}
\label{sec:image_analysis}

The stacked image in Fig.~\ref{fig:stacked} clearly shows the four
bright rings identified in the discovery ATel \citep{beardmore:15} and
the five rings identified in \citet{vasilopoulos:16}.  In addition to
the previously identified rings, the {\em Chandra} and the stacked
{\em Swift} images show a number of fainter rings.  In total, we
identify {\em eight} separate rings in the {\em Swift} image (which
has more complete spatial coverage than the {\em Chandra} image),
labeled [a]-[h] from inside out, indicated in location and radial
extent by white dots and characteristic inner and outer ring-radii,
respectively, in Fig.~\ref{fig:stacked}. Rings [c], [f], and [h] were
not included in the analysis by \citep{vasilopoulos:16} because they
are not apparent in individual {\em Swift} pointings --- detection
from the {\em Chandra} image and the stacked {\em Swift} image, as
well as from the radial intensity profiles discussed in
\ref{sec:profiles}, is straightforward, however. The five rings
discussed in \citet{vasilopoulos:16} correspond to the rings [a], [b],
[d], [e], and [g].

While the outskirts of the image are noisy, the image shows clearly
that the rings are {\em not} azimuthally uniform.  As we will discuss
below, rings [a] and [b] also have significant {\em radial}
sub-structure in the {\em Chandra} images.

A color gradient is apparent in Fig.~\ref{fig:stacked}, with bluer
regions toward the Galactic plane (in the direction of position angle
$55^{\circ}$, as expected for a gradient in column density and thus
photo-electric absorption, toward the plane. We will discuss the
asymmetry of the column density distribution in the field of view
(FOV) relative to the position of V404 Cyg and its implications for
image- and spectral analysis in \S\ref{sec:NH}.

\subsection{Radial Intensity Profiles and Time of the Peak Flare}
\label{sec:profiles}

Because the rings are concentric about the position of V404 Cyg, it is
appropriate to construct radial intensity profiles.  Following the
analysis of the Circinus X-1 light echo in \citet{heinz:15}, we
generated azimuthally averaged radial surface-brightness profiles for
each observation centered on the (attitude-corrected) position of V404
Cyg.  Profiles are computed on a logarithmic radial grid to allow for
deconvolution with the annular profile of the light echo generated by
a single thin dust sheet, which can be calculated from the soft X-ray
lightcurve of the outburst \citep[see][for details]{heinz:15}.

\begin{figure}[t]
  \center\resizebox{\columnwidth}{!}{\includegraphics{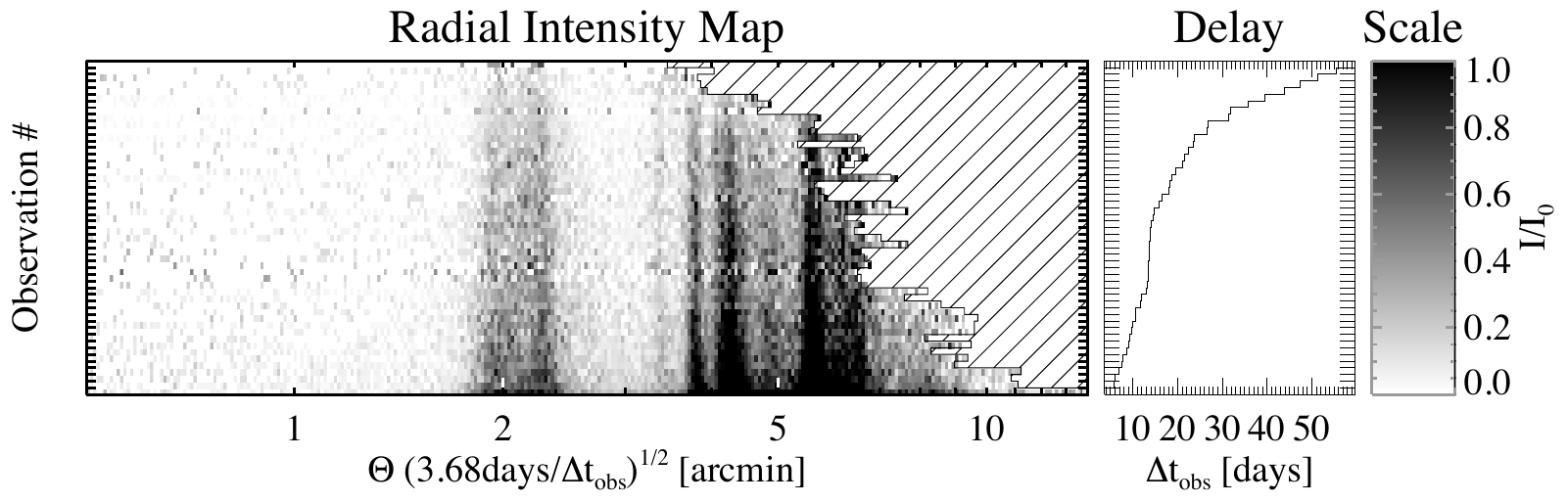}}
  \caption{{\em From left to right:} (1) Map of the radial intensity
    profiles
    $I_{1-2}\left(1-x\right)^2\left(\Theta_{\rm
        sc}/1000''\right)^{\alpha}$
    of all {\em Swift} images used in this analysis in the 1-2 keV
    band (with $\alpha=3.0$ from \S\ref{sec:fit_results}), as a
    function of observation number and angular distance
    $\Theta \left(\Delta t_{obs 1}/\Delta t\right)^{1/2}$ from \vcyg ~
    (the scaling factors remove the time dependence of the angular and
    intensity scale to lowest order). The hatched area shows the edge
    of the field of view. Vertical features correspond to individual
    clouds. (2) Delay time between main flare and {\em Swift}
    observation as a function of observation number. (3) Linear
    intensity scale used in the leftmost panel.\label{fig:profiles}}
\end{figure}

All 1-2 keV {\em Swift} profiles are shown in the left panel of
Fig.~\ref{fig:profiles} as a function of observation number and
angular distance from V404 Cyg.  Because the angular scale of the echo
increases with time as $\theta \propto \Delta t^{1/2}$ according to
eq.~\ref{eq:theta}, the radial bins for the intensity profiles were
chosen to increase by the same ratio, that is, the profiles were
extracted in 300 logarithmically spaced bins between
$\theta_{\rm min}=0'.5\,\sqrt{\Delta t/\Delta t_{0}}$ and
$\theta_{\rm max}=14'\,\sqrt{\Delta t/\Delta t_{0}}$ such that the
echo of the main flare on MJD57199.8 appears stationary on the grid of
intensity profiles{\color{black}{, displayed against the angular scale
    $\theta_{0}$ of ObsID 00031403071}}.  The success of this method
in capturing the main features of the echo can be seen from the
vertical alignment of the echo from each ring in
Fig.~\ref{fig:profiles}.

{\em Chandra} angular profiles were extracted on a finer angular grid
than the {\em Swift} data to take advantage of the better angular
resolution (and thus not over-plotted in Fig.~\ref{fig:profiles}).

Using eq.~(\ref{eq:intensity}), the intensity scale of each row in the
image was adjusted by a factor of
$(1 - x)^2\left(\theta_{\rm sc}/1000''\right)^{\alpha}$ [where we use
$\alpha \sim 3$ from \S\ref{sec:fit_results} and where the scattering
angle is calculated using eq.~(\ref{eq:theta_scatter}), assuming the
emission is dominated by the echo from the bright flare on MJD
57199.8] to remove (to lowest order) the strong temporal decline in
echo brightness due to the increase in scattering angle, allowing us
to show all profiles in a single image.

We analyzed the data in four energy bands, 0.5-1, 1-2, 2-3, and 3-5
keV.  Profiles were background subtracted by constructing a clean sky
and instrumental background level for the V404 Cyg field of view from
the pre-flare XRT imaging observations.

In order to explore the possibility of multiple flares undetected due
to gaps in coverage, and in order to determine the time of the main
flare, we plotted the cross-correlation of the radial profiles of the
two {\em Chandra} data sets in {\color{black}{the third panel from the
    top}} of Fig.~\ref{fig:lightcurve} against the time of the flare
(calculated from the time lag). The cross-correlation shows a clear
peak on MJD 57199.8, indicating that the main flare that gave rise to
the rings does indeed correspond to the main flare seen by {\em
  INTEGRAL}, consistent with the findings in
\citet{vasilopoulos:16}. The figure also shows the cross-correlation
of the intensity profiles of {\em Swift} ObsID 0031403071 and
0031403079, which shows the most well-defined peak of all the {\em
  Swift} cross-correlations (all other cross-correlations of {\em
  Swift} profiles are less well defined).  The peak is consistent with
the peak derived from the {\em Chandra} data, though with somewhat
broader wings.

\subsection{The Width of Rings [a]-[d]}
\label{sec:psf}

The innermost two rings [a] and [b] overlap significantly in the {\em
  Swift} image, but are clearly resolved in both of the {\em Chanda}
images. It is also clear that the rings are not azimuthally symmetric,
both in color and total intensity in all images, which will be
discussed further in \S\ref{sec:NH}.

Figure~\ref{fig:chandra_image} shows that several of the rings are
very sharp. In particular, the two innermost rings [a] and [b] contain
bright, sharp arclets in the Southern and North-Western halfs of the
rings, respectively, denoted by arrow-marks in the figure.  Rings [c]
and [d] also appear sharp, while the outer rings [e] and [f] appear
broader in Fig.~\ref{fig:chandra_image}.  We identify the sharp
brightness peaks with the echo from the main flare at the end of the
outburst on MJD 57199.8.  Emission from {\em earlier} sub-flares of
the outburst will be located at {\em larger} angles according to
eq.~(\ref{eq:theta}).

\begin{figure}[t]
  \center\resizebox{\columnwidth}{!}{\includegraphics{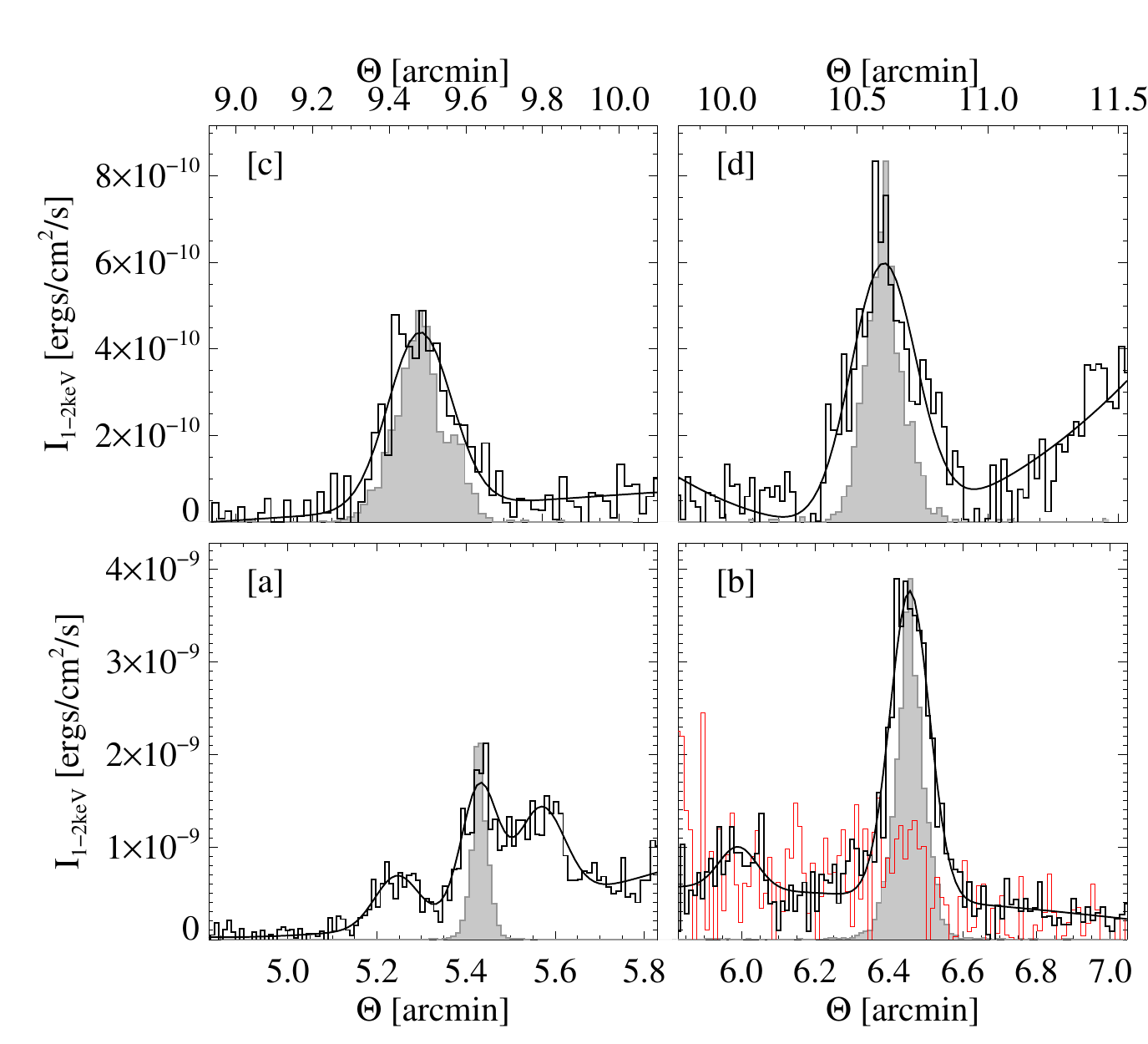}}
  \caption{Radial intensity profiles across rings [a]-[d] (taken at
    position angles $90^{\circ} - 270^{\circ}$,
    $9.5^{\circ} - 72^{\circ}$, $90^{\circ} - 270^{\circ}$, and
    $90^{\circ}-270^{\circ}$, respectively) in the 1-2 keV band from
    {\em Chandra} obsID 17704 (black histograms). Overplotted are
    Gaussian fits to the ring components (black curves) and radial
    profiles of the 1-2 keV {\em Chandra} PSF at the location of rings
    (filled gray histograms). For comparison, the red histogram in
    panel [b] shows the radial profile for ring [b] taken at position
    angles $167^{\circ}-333^{\circ}$.}\label{fig:psf}
\end{figure}

Figure \ref{fig:psf} shows the radial 1-2 keV intensity profile across
the brightest segments of rings [a]-[d] in Chandra ObsID 17704. Ring
[a] was extracted between position angles $90^{\circ} - 270^{\circ}$,
ring [b] over position angles $9.5^{\circ}$ and $72^{\circ}$, and
rings [c] and [d] over the entire region covered by the chips of the
ACIS S array, roughly from position angle $90^{\circ}$ to
$270^{\circ}$.

Because of the large angular size of the rings (5'.5, 6'.44, 9'.5, and
10'.6, respectively), the {\em Chandra} point-spread-function (PSF) is
significantly wider than near the aim point, and a by-eye inspection
of the width of the rings is insufficient to determine whether the
rings are resolved or not.

In order to evaluate whether the rings are resolved, we generated
matched {\em Chandra} PSFs using {\tt CHART/MARX} at the positions of
the peak intensities of the rings, using the aspect file for ObsID
17704. The simulated PSFs were reduced identically to ObsID 17704;
radial surface-brightness profiles of the PSFs at the radii of rings
[a]-[d] are over-plotted in gray in Fig.~\ref{fig:psf} on the narrowest
and highest peak of the respective ring. All rings are marginally
resolved.  Ring [a] shows clear sub-structure with at least three
sub-peaks, of which we chose the narrowest for comparison.

From equation \ref{eq:theta}, the radial extent of each ring is
determined by {\em both} the line-of-sight extent $\Delta x$ of the
cloud responsible for the ring {\em and} the structure of the
light curve (i.e., the temporal width of the flare responsible for the
ring).

The sharpness of the rings implies very concentrated dust clouds
responsible for the rings and a very sharp peak in the flare
lightcurve.  Given the gaps in the lightcurve coverage, we cannot
distinguish whether the width is due to spatial cloud extent or
duration of the flare. However, we can place upper limits on the FWHM
line-of-sight extent of the four clouds [a]-[d]. These are listed in
Table \ref{tab:psf}, along with the inferred distances to the clouds.

In particular, we find an upper limit on the line-of-sight extent of
cloud [b] of $D_{\rm [b]} \lesssim 8.8\,{\rm pc}$.  This is comparable
to the transverse size scale of $L_{\rm [b]} \sim 5\,{\rm pc}$ for
cloud [b] inferred from the angular size of the brightest ringlet,
which is roughly 7 arcminutes across, and, since the clouds
responsible for rings [a]-[h] have roughly comparable column density,
the extent along the line of sight is comparable to the expected
typical cloud size toward V404 Cyg in general.

\subsection{The Non Axi-Symmetry of Rings [a] and [b]}
\label{sec:NH}

As was found in \citet{heinz:15}, rings from dust scattering echoes
can show significant deviations from axi-symmetry due to variations in
the column density of the scattering dust concentrations.
{\color{black}{Asymmetries have also been found in a number of dust
    scattering halos by \citet{seward:13}, \citet{valencic:15} and
    \citet{mccollough:13}.}}

For typical {\em physical} clump sizes of order a few parsec,
appropriate for moderate-mass clouds with hydrogen column densities of
order $10^{21}\,{\rm cm^{-2}}$ \citep[e.g.][]{larson:81,heyer:09}, the
typical {\em angular} scale of clouds between us and V404 Cyg will be
of the order of a few arcminutes to tens of arcminutes and thus
potentially {\em smaller} than the FOV covered by the echo.  For
random placement of clouds, we should therefore {\em expect} that the
different clouds responsible for the rings (a) will not be centered on
the position of V404 Cyg, and (b) may show significant density
gradients across the FOV. Furthermore, since the source is about 2
degrees south of the Galactic plane, we would expect cloud centroids
to be preferentially off-set to the North-West of V404 Cyg, in the
direction toward the Galactic plane.

This should induce a deviation from axi-symmetry in the brightness
distributions of the rings, and it should also lead to non-monotonic
radial brightness evolution as the rings expand, complicating analysis
of the dust scattering signal, because the column density of each
cloud cannot necessarily be considered uniform across the image.

Of all rings visible in Figs.~\ref{fig:chandra_image} and
\ref{fig:stacked}, ring [b] shows the clearest signs of deviation from
axi-symmetry, with the North-West section of the ring being much
brighter than the other three sections.  The spectrum of the
North-West section of ring [b] is also much harder, supporting the
notion that cloud [b] has a much higher column density in the
North-West direction, which increases {\em both} the scattering
intensity {\em and} the amount of photo-electric absorption.

{\color{black}{Because the photons for {\em each} scattering ring must
    pass through {\em all} clouds along the line of sight (either
    before or after scattering), the azimuthal variation in the column
    in each cloud will be imprinted on all scattering rings. Thus,
    some of the general trend of the color in Fig.~\ref{fig:stacked}
    can be attributed to cloud [b] alone. The compounded azimuthal
    asymmetry of the absorption column must be included when modeling
    the emission of the rings and will be described in more detail in
    \S\ref{sec:spectral_model}.}}

The bottom-right panel of Fig.~\ref{fig:psf} shows a comparison of the
1-2 keV angular profiles of ring [b] from the North-West quadrant
(black curve) to the Eastern half of the ring (red curve), indicating
that the North-Western part is about four times brighter than the
Eastern part of the ring.

We quantified this by simple parametric spectral fits to the four
quadrants of rings [a] and [b] of Chandra ObsID 17704.  We extracted
spectra of the rings in annular sections in the North-West,
South-West, South-East, and North-East quadrants and fit them with a
simple absorbed powerlaw model
{\color{black}{(${\tt phab}\cdot{\tt powerlaw}$}} in {\tt XSPEC}),
where the powerlaw model, written in photon flux $\phi_{E}$, takes the
form
\begin{equation}
  \phi_{E}=\phi_{0}\left(\frac{E}{1\rm keV}\right)^{-\Gamma}
\end{equation}
We tied the powerlaw slope of all spectra across all rings but left
the photo-electric column $N_{\rm H}$ and the powerlaw normalization
for each ring as free parameters.  The resulting fit parameters and
uncertainties are listed in Table~\ref{tab:17704_table}.

We find that, for ring [b], the normalization of the powerlaw (i.e.,
the scattered intensity after removing the effects of photo-electric
absorption) is a factor of 4 larger than in the other three quadrants,
a highly significant deviation, while ring [a] is brightest in the
South-Western quadrant.  The photo-electric neutral Hydrogen column
density in the North-West quadrant for both rings [a] and [b] is
larger by about $3\times\,10^{21}\,cm^{-2}$ than in the other three
quadrants.  Since ring [a] is not brighter in this direction, we
attribute the increase in column density primarily to ring [b], though
it must be kept in mind that the total photo-electric column is
expected to increase toward the Galactic plane in the North-West
direction, and therefore the total column of clouds [c]-[h] will
likely, on average, be larger in the North-West quadrant.

This is clear evidence that the centroid of cloud [b] is off-set from
the direction of V404 Cyg and that the dust scattering column density
{\em cannot} be assumed uniform across the image.  By extension, we
cannot make the assumption of uniform column density for {\em any} of
the clouds in our analysis of the dust scattering echo.

\begin{figure}[t]
  \center\resizebox{\columnwidth}{!}{\includegraphics{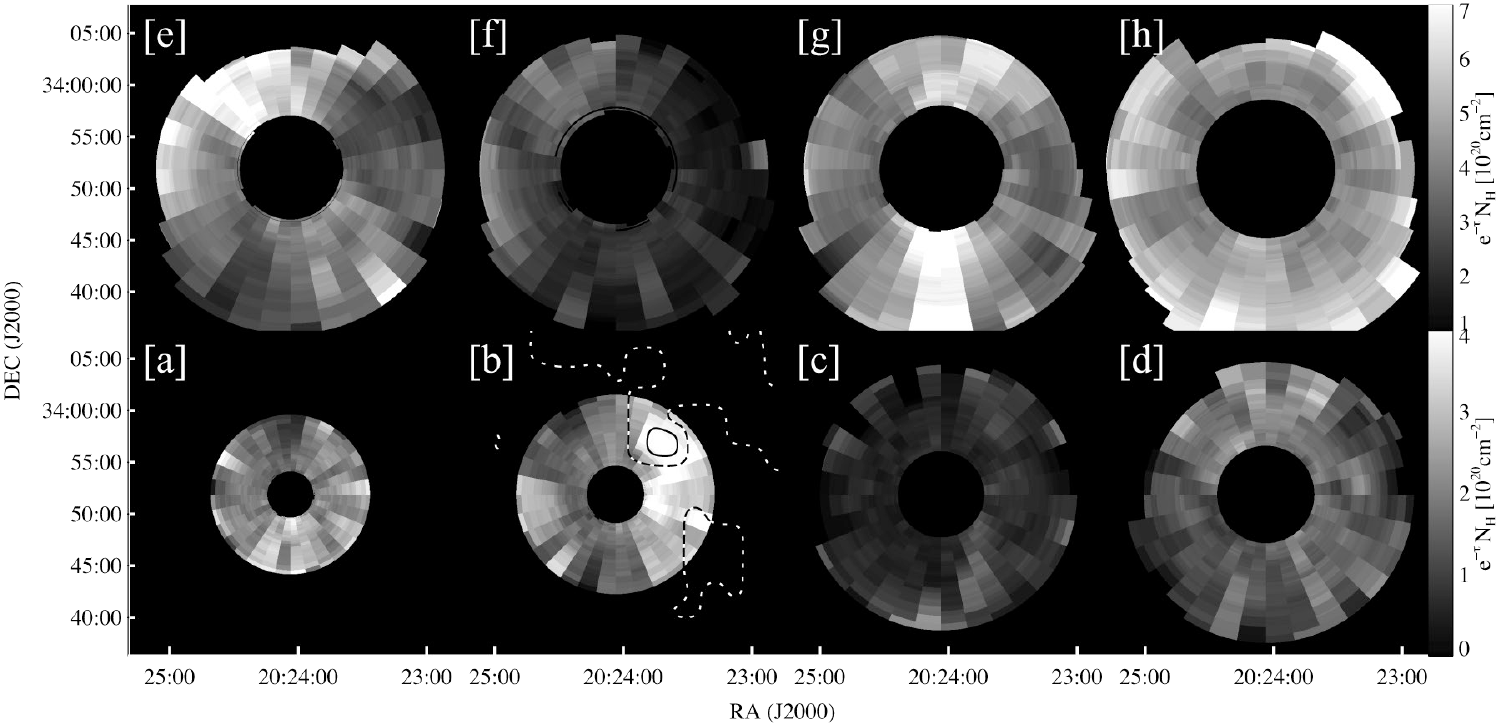}}
  \caption{Reconstructed cloud images in the 1-2 keV band, calculated
    from stacking all {\em Swift} images for each individual ring
    (using the radial bins indicated in Fig.~\ref{fig:stacked} for
    each ring) and binning them into 32 azimuthal sections. The image
    is scaled by
    $x(1-x)D/({\mathcal F}\Delta \varphi\,c\,d\sigma/d\Omega)$ to show
    the absorbed column density $e^{-\tau}N_{\rm H}$ according to
    eq.~(\ref{eq:fluence}), where ${\mathcal F}$ and $d\sigma/d\Omega$
    are taken from the best fit spectral model discussed in
    \S\ref{sec:fit_results} and
    $\Delta \varphi=11.25^{\circ}=0.196\,{\rm rad}$ is the angular bin
    size. The images show clear column density variations across the
    field for each cloud. Overlaid on panel [b] are contours of excess
    extinction $\Delta E(B-V)=0.4$ (dashed) and $\Delta E(B-V)=0.8$
    (solid) in the 2-2.5 kpc distance range from Pan-STARRS,
    coincident with the peak in column density of cloud
    [b].}\label{fig:cloud_images}
\end{figure}

To test the azimuthal uniformity of the different clouds, we generated
stacked cloud images of all {\em Swift} observations in the 1-2 keV
band, where the signal-to-noise is highest. The result is shown in
Fig.~\ref{fig:cloud_images}. 

To generate the cloud images, we divided each {\em XRT} image into
eight rings (denoted by the white marks in Fig.~\ref{fig:stacked}
{\color{black}{in units of $\theta_{0}$, adjusted to the angular scale
    of the observation using eq.~\ref{eq:theta}, that is, the ring
    extraction is performed in sky coordinates $\theta$, not in scaled
    coordinates $\theta_{0}$),}} and each ring into 32 azimuthal bins,
and derived the 1-2 keV flux of each bin from the total exposure
corrected count rate.

Following eq.~(\ref{eq:fluence}), the flux in each bin was multiplied
by a factor
$x(1-x)D/[\Delta \varphi c {\mathcal F} (d\sigma/d\Omega)]$ to remove,
to lowest order, the dependence on cross section and cloud distance
$x$. Here, $\Delta \varphi = 11.25^{\circ}=0.196\,{\rm rad}$ is the
angular size of each bin.  We used best-fit values for the cross
section and fluence from our fiducial MRN1 spectral model listed in
\S\ref{sec:fit_results}.  Each bin therefore contains a local measure
of $e^{-\tau_{\rm ph}}N_{\rm H}$, i.e., not corrected for
photo-electric absorption.  {\color{black}{Note again that {\em each}
    cloud affects {\em all} rings by photo-electric absorption at an
    angular scale set by the relative distances to the cloud
    responsible for the scattering and that responsible for the
    absorption. The compounded effect of photo-electric absorption is
    treated in detail in \S\ref{sec:spectral_model}.}} All 50 images
for a given ring were then stacked --- at the actual physical angular
scale $\theta$ rather than scaled to $theta_{0}$ --- to produce a
single dust map of the cloud generating the ring.

The outer regions of the images are noisy, but there is clear evidence
of non-uniformity in all of the cloud images.  In particular, cloud
[b] appears to have a strong local peak in the North-West quadrant
centered on location {\color{black}{(RA=20:23:40.9,DEC=33:56:39).}}

\begin{figure}[t]
  \center\resizebox{\columnwidth}{!}
  {\includegraphics{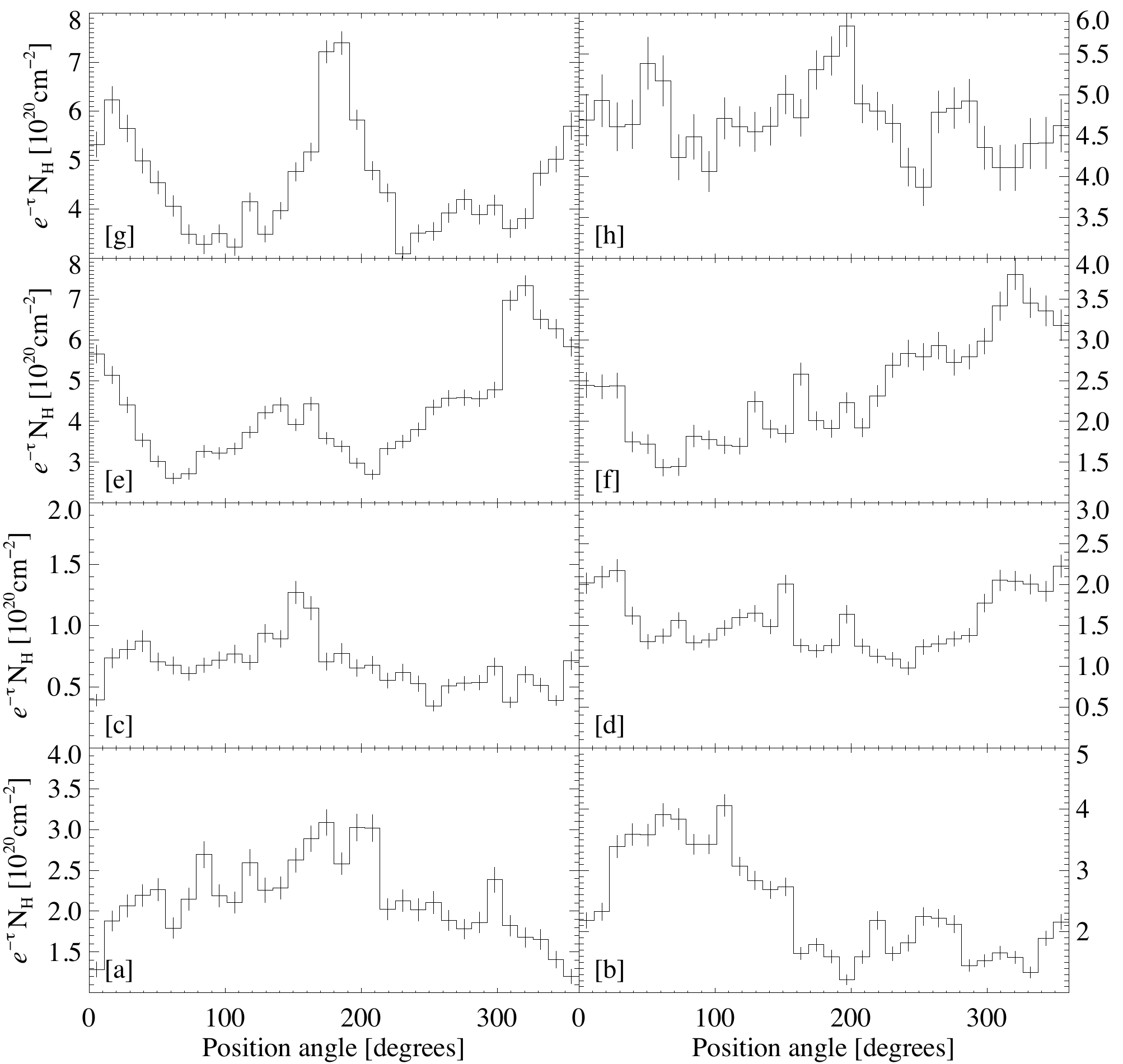}}
  \caption{Angular column-density profiles constructed by summing each
    radial bin in Fig.~\ref{fig:cloud_images} in radius.  Error bars
    indicate 1-sigma Poisson errors in each
    bin.}\label{fig:angular_profiles}
\end{figure}

We tested the significance of the azimuthal variations by summing all
angular profiles in the 1-2 keV band used to construct
Fig.~\ref{fig:cloud_images} along the radial direction for each cloud,
thus creating 1D azimuthal intensity profiles, shown in
Fig.~\ref{fig:angular_profiles}.

The azimuthal intensity peaks in rings [a], [b], [d], [e], [f], and
[g] are clearly visible in Fig.~\ref{fig:angular_profiles}. The
measured variance in the profiles is larger than the mean error in the
intensity by factors of 3.4, 5.6, 2.1, 5.0,10.1,5.4,5.3, and 2.6 from
ring [a] to [h], respectively, indicating that the angular variations
seen in the profiles are significant. Note that error bars indicate
the 1-sigma Poisson uncertainties in each angular bin.

\subsection{Reconstruction of the Soft X-ray Lightcurve of the
  Outburst from the {\em Chandra} Light Echo}
\label{sec:lightcurve}

In \S\ref{sec:deconvolution}, we will employ the same
radial-deconvolution technique of the ring profile used in
\citet{heinz:15} to recover the dust distribution along the line of
sight.  That is, we use the fact that for a thin scattering sheet of
dust the intensity profile of the light echo is described by
eq.~(\ref{eq:intensity}) for a single $x$ and use eq.~\ref{eq:theta}
to relate $\Delta t$ to $\theta$.

If the complete light curve $F(\Delta t)$ of the outburst is known,
and we make the assumption that the scattering cross section can be
reasonably approximated as a powerlaw in scattering angle
$d\sigma/d\Omega \propto \theta_{\rm sc}^{-\alpha}$ over a moderate
range in scattering angle, the radial intensity profile can then be
decomposed into dust echoes from a series of scattering screens along
the line of sight, using the kernel function
\begin{equation}
  K(z)=F(\Delta t_{\rm flare}z^2)z^{-\alpha}
  \label{eq:kernel}
\end{equation}
where $z\equiv \theta/\theta_{\rm flare}$ and $\theta_{\rm
  flare}=\theta(\Delta t_{\rm flare})$ from eq.~(\ref{eq:theta}),
using $\Delta t_{\rm flare} \equiv t_{\rm obs} - t_{\rm flare}$.  Such
a deconvolution will yield a distribution of scattering depth as a
function of fractional dust distance $x$:
\begin{equation}
  \Delta \tau(x) = \frac{\Delta N_{\rm H}}{\left(1-x\right)^2}\frac{d\sigma}{d\Omega}
\end{equation}
The deconvolution is computationally straight forward if the radial
profiles are extracted on logarithmic bins.  We employ the
Lucy-Richardson maximum-likelihood deconvolution
\citep{richardson:72,lucy:74} implemented in the {\tt IDL ASTROLIB}
library.

In the case of the V404 Cyg outburst, the gaps in coverage of the
light curve (and the fact that no soft-X-ray telescope observed the
outburst frequently enough to provide a reliable 1-2 keV light curve)
do not allow the direct application of this technique. However, we can
use the high angular resolution of {\em Chandra} and the
non-axisymmetric brightness profile of ring [b] to reconstruct the
light curve of the V404 Cyg outburst.  We will then use this
reconstructed light curve in our deconvolution of the echo.

From the profile shown in Fig.~\ref{fig:psf}b and the spectral
extraction discussed in \S\ref{sec:NH}, we know that the column
density toward cloud [b] is much larger in the North-West quadrant
than in any other quadrant, while cloud [a] and any intervening clouds
between cloud [a] and [b] (which may contribute to ring emission in
the annular range between 5' and 7'), are not strongly concentrated in
the North-West quadrant.

We also know from the image and analysis of the intensity profiles
that the annuli outside of ring [b] between 7' and 9'.2 do not show
evidence of ring emission.  Thus, the emission in the North-West
quadrant from about 6'.2 outward is dominated by emission from ring
[b]. We can therefore use the {\em difference} in intensity between
the North-West quadrants and the Eastern quadrants to {\em isolate}
emission from cloud [b] only.

To lowest order, this difference will be the emission produced by the
excess in column density of cloud [b] in the North-Western quadrant
and will correspond the the light echo of a single, thin cloud of
dust.  The bright peak of ring [b]$_{\rm NW}$ is clearly due to the
main flare of the outburst. We expect lower intensity emission from
prior sub-flares of the outburst to lie in the angular range from
6'.5-7'.4.

In order to extract this scattering response, we employed an iterative
difference procedure between the Eastern and North-Western intensity
profiles, isolating the difference in profiles from 6'.3 outward to
construct the dust scattering kernel (that is, the intensity profile
produced by cloud [b] only).

We first constructed a zero-order kernel by deconvolving the
North-West profile of ring [b] with the Chandra PSF for ring [b]
discussed in \S\ref{sec:psf} and selected only emission outward of
6'.3 (that is, we reject any difference in the intensity profiles
inward of the brightest peak in the intensity profile of ring [b],
which we attribute to the flare at the end of the outburst, such that
any additional emission from ring [b] must lie at larger angles).  We
re-convolved this profile with the Chandra PSF to derive the
zero-order kernel function $K(z)_{0}$, using $\alpha\sim 3.0$ for our
best-fit standard MRN1 dust scattering cross section from
\S\ref{sec:fit_results}\footnote{We have verified that all deconvolutions
used in this paper are insensitive to the exact choice of $\alpha$ in
a range from $2.7 < \alpha < 4.0$.}.

We then deconvolved the {\em Eastern} intensity profile in the angular
range 5'.0-7'.4 with $K(z)_{0}$, isolated the deconvolution to
contributions outward of 6'.3 (setting the interior to zero),
re-convolved with $K(z)_{0}$ and subtracted the result from $K(z)_0$
to derive the next iteration in the kernel, $K(z)_{1}$.  We repeated
the procedure until the total absolute fractional difference between
two iterative kernel functions $K(z)_n$ and $K(z)_{n+1}$ was less than
$10^{-4}$ (convergence was reached after 5 iterations).

This procedure self-consistently accounts for the emission both from
ring [b] as well as contamination from dust responsible for emission
interior to ring [b] (which will contribute to the total intensity
outward of ring [b] given the length of the outburst).

From this self-consistently derived kernel, we can reconstruct the
X-ray light curve by inverting eq.~(\ref{eq:kernel}).  The resulting
light curve is plotted in the second top-most panel in
Fig.~\ref{fig:lightcurve}, both in its deconvolved form and convolved
with the {\em Chandra} PSF.  We used the latter for analysis. 

The statistical uncertainties plotted in the figure were calculated
using Monte-Carlo simulations by randomizing the intensity profiles
with the appropriate amount of Poisson noise and repeating the
iterative deconvolution procedure for each of the 1000 realizations.

The re-constructed light curve shows a clear peak at MJD 57199.8,
within 0.2 days of the main flare of the outburst identified by
{\color{black}{{\em INTEGRAL} (bottom two panels) and the
    cross-correlations (third panel from the top).}}  The
deconvolution suggests that the flare may have had two peaks, but
following the discussion in \S\ref{sec:psf}, we cannot distinguish
whether the width of ring [b] is due to the properties of the flare or
the line-of-sight distribution of cloud [b]. The smooth curve
indicates that the temporal resolution of our iterative reconstruction
is about a half day, and any claims for a double-peaked lightcurve on
timescales shorter than that would be speculative at best.

The reconstructed light curve shows a broad tail of emission between
MJD 57192 and 57199, consistent with the level of activity seen in the
hard X-rays by both {\em Swift} BAT and {\em INTEGRAL} JEMX.  The lack
of any significant peaks in the tail suggests that no major flares
were missed despite the gaps in coverage.

The cumulative fluence of the {\em Chandra}-reconstructed light curve
(plotted in the top panel of Fig.~\ref{fig:lightcurve}) is consistent
with the fluence distribution of the {\em INTEGRAL} lightcurve. Both
suggest that about 70\% of the total outburst fluence were
concentrated in the major flare on MJD 57199.8.  Using a single flare
time and analyzing the echo based only on the major flare on MJD
57199.8 would thus {\em miss} about a third of the fluence in the
roughly week-long precursor to the main flare.  The emission-weighted
mean precursor time is $t_{\rm pre} = $ MJD 57195.3.

\subsection{The Dust Distribution Toward V404 Cyg}
\label{sec:deconvolution}

\begin{figure}[t]
  \center\resizebox{\columnwidth}{!}{\includegraphics{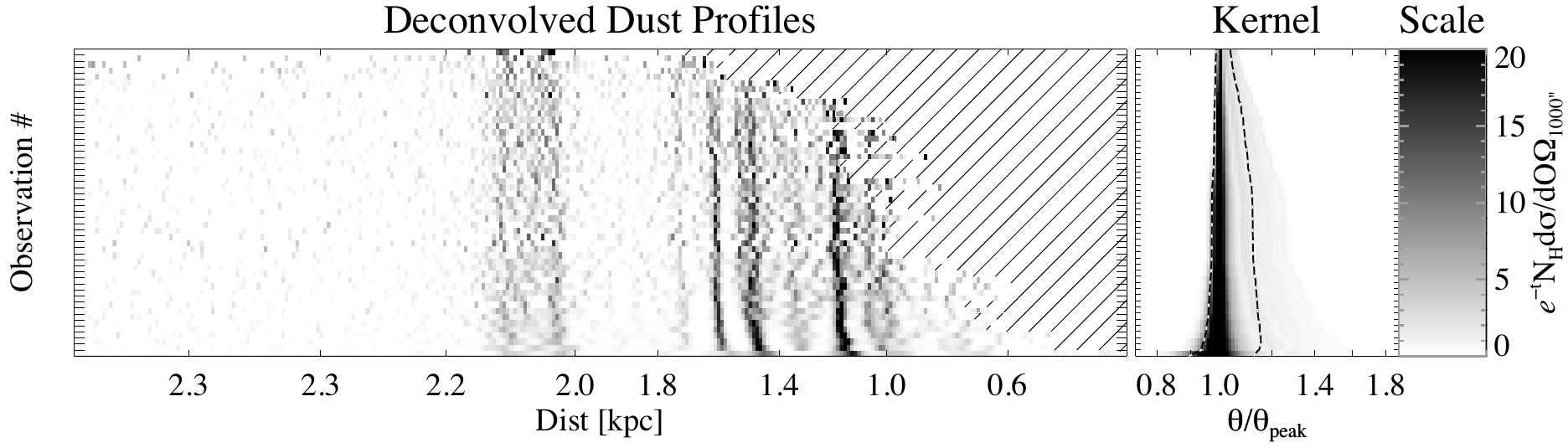}}
  \caption{{\em From left to right:} (1) Map of deconvolved intensity
    profiles as a function of dust distance; (2) deconvolution kernel
    used to derive panel (a) from Fig.~\ref{fig:profiles}, calculated
    using eq.~\ref{eq:kernel} and the reconstructed outburst
    lightcurve from Fig.~\ref{fig:lightcurve}. The dashed lines
    indicate the 10\% and 90\% flux contours, i.e., 80\% of the kernel
    flux is contained in the area between the lines; {\color{black}{(3)
        intensity scale bar for panel (1).}}}\label{fig:deconvolution}
\end{figure}

With a reconstructed outburst lightcurve in hand, we can employ the
deconvolution procedure used in \citet{heinz:15} and outlined in
\S\ref{sec:lightcurve} to derive the dust distribution toward V404
Cyg.  The kernel defined in eq.~\ref{eq:kernel} used for this
deconvolution is a function of time: For earlier observations, the
range in angle spanned by the kernel is larger, that is, the rings
from earlier observations are more spread out.  This is visible in
Fig.~\ref{fig:profiles}, which clearly shows the rings narrowing with
time (relatively to their angular siez).  For a total outburst
duration of $\delta t_{\rm burst} \sim 9\,{\rm days}$, the angular
spread at late observing times evolves roughly as
\begin{equation}
  \frac{\Delta \theta}{\theta} \sim \sqrt{\frac{\delta t_{\rm
        burst}}{\Delta t_{\rm obs}}}.
  \label{eq:thinning}
\end{equation}
This effect can be seen in the vertical ``thinning'' of the intensity
profiles in the left panel of Fig.~\ref{fig:profiles} and of the outer
envelope of the kernel function plotted in the right panel of
Fig.~\ref{fig:deconvolution}.  The dashed contour in
Fig.~\ref{fig:deconvolution} shows the region of the kernel containing
80\% of the kernel flux.  Note that this contour does not exactly
track the thinning described in eq.~(\ref{eq:thinning}) because the
part of the kernel at larger scattering angles become brighter with
time relative to the part of the kernel at smaller angles as the range
in scattering angles spanned by the kernel decreases.

Each radial profile is deconvolved with the appropriate kernel.  In
order to correct for the dependence of intensity on $x$ and
$\theta_{\rm sc}$ (see eq.~\ref{eq:intensity}), we multiply each
deconvolution by
$(1-x)^2\left(\theta_{\rm sc}/1000''\right)^{\alpha}$, where $x$ and
$\theta_{\rm sc}$ for each angular bin are calculated from equations
(\ref{eq:x}) and (\ref{eq:theta}), respectively, and mapped to the
corresponding distance $D=xD_{\rm V404}$, and we use
$\alpha \sim 3.0$.  The resulting curves measure the scattering depth
$\Delta\tau_{\rm sc} = e^{-\tau_{\rm ph}}\Delta N_{\rm
  dust}\frac{d\sigma}{d\Omega}(1000'')$
and are thus proportional to the dust column density, not corrected
for photo-electric absorption.  The resulting scattering depths of all
{\em Swift} observations are plotted in the left panel of
Fig.~\ref{fig:deconvolution}.

We will derive properly un-absorbed column densities of the different
clouds in \S\ref{sec:fit_results}.  However, all of the rings are
affected by absorption in a similar fashion: the total column
affecting each ring is just the sum of the column densities of all
clouds. Moderate differences in the total absorption will arise from
the azimuthal variations in each ring.  The derived deconvolutions
then allow us to derive the {\em relative} distribution of dust along
the line of sight.

\begin{figure}[t]
  \center\resizebox{\columnwidth}{!}{\includegraphics{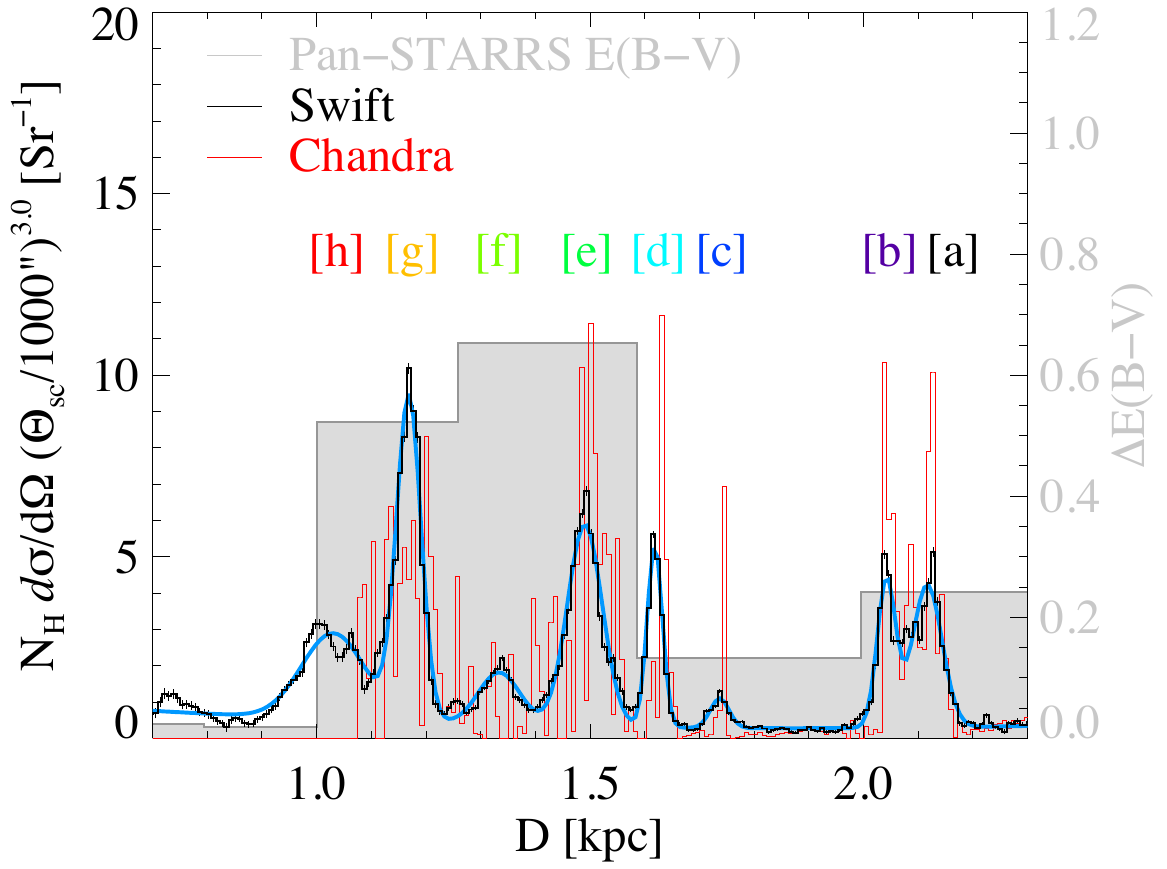}}
  \caption{Dust distribution along the line-of-sight toward V404 Cyg,
    plotted as scattering depth corrected
    $\tau_{\rm eff} \equiv N_{\rm H} d\sigma/d\Omega \left(\Theta_{\rm
        sc}/1000"\right)^{\alpha}$,
    derived from the deconvolutions of the {\rm Chandra} ObsID 17704
    (red) and {\em Swift} (black) intensity profiles, plotted as a
    function of dust distance $D$ from Earth. The {\em Swift}
    distribution is the weighted average over all {\em Swift}
    observations (i.e., the weighted sum along the vertical column of
    Fig.~\ref{fig:deconvolution}). The solid blue line shows the
    Gaussian fit to the {\em Swift} data (black histogram) for eight
    distinct dust concentrations. Overlaid in gray is the differential
    extinction distribution in the direction of V404 from the public
    Pan-STARRS $E(B-V)$ data, with $\Delta E(B-V) \propto N_{\rm H}$,
    showing general agreement with the dust distribution derived from
    the X-ray light echo.}\label{fig:column}
\end{figure}

To derive a high signal-to-noise 1D dust distribution along the
line-of-sight, we performed a weighted average of the dust
distributions from each {\em Swift} observation, plotted as a black
histogram in Fig.~\ref{fig:column}.  Over-plotted in red is the dust
distribution derived from {\em Chandra} ObsID 17704.

The histograms show at least eight separate clouds (as identified in
Fig.~\ref{fig:stacked}), labeled [a]-[h], following the above naming
convention, confirming the by-eye identifications in the stacked image
in Fig.~\ref{fig:stacked}.  The positions and scattering depths of the
clouds derived from {\em Swift} and {\em Chandra} are consistent with
each other out to the edge of the {\em Chandra} FOV (at distances
larger than 1.2 kpc).  Clouds [a] and [b] show higher column in the
{\em Chandra} deconvolution, which we attribute to the fact that cloud
centroids are clearly offset from the position of V404 Cyg (as
discussed in \S\ref{sec:NH}): the {\em Chandra} observation was taken
late during the echo, with rings [a] and [b] intersecting the cloud
centroid, while the weighted average {\em Swift} dust distribution
contains earlier, high signal-to-noise observations that are expected
to have lower mean $N_{\rm H}$ for rings [a] and [b].

For further analysis, we fitted each of the eight clouds with a
Gaussian, over-plotted in blue in Fig.~\ref{fig:column}.  As expected
from the higher angular resolution of the {\em Chandra} maps, the
peaks in the dust distribution derived from ObsID 17704 are narrower.
The {\em Chandra} profile (see Fig.~\ref{fig:psf}) also suggests the
presence of additional rings between [a] and [b], which are not
discernible in the {\em Swift} profile.  We label these rings as ring
[a.2] and [b.1], while we label the main ring as rings [a.1] and
[b.2].  In Table \ref{tab:clouds} we list the positions and
line-of-sight widths inferred from the Gaussian fits.  Where rings are
sufficiently covered by {\em Chandra}, we list the {\em Chandra}
values for their higher precision in distance and depth and split
rings [a] and [b] into separate components, determined from {\em
  Chandra} ObsID 17704.  The table also lists the mean and the
variance of the cloud column densities, as well as parameters of the
grain size distribution, determined from spectral fits described in
\S\ref{sec:spectra} and \S\ref{sec:discussion}.

For reference, we overplot the dust distribution from the
$\Delta E(B-V)$ extinction map from Pan-STARRS in gray
\citep{green:15}.  The $E(B-V)$ spatial resolution is insufficient to
map each cloud to an individual extinction peak, but the general
distribution is consistent, with rings [e]-[h] corresponding to the
majority of the dust, and a clear peak at the positions of rings [a]
and [b]. The low value of $E(B-V) < 0.1$ short-ward of 1 kpc and the
lack of any jumps in extinction in that distance range suggests that
clouds [a]-[h] contain essentially all of the dust between us and V404
Cyg (see also the discussion in \S\ref{sec:fit_results}).

 Clearly, the power of an extensive series of observations as provided
by {\em Swift} lies in the ability to probe both the dust distribution
and the dust scattering cross section over a large range in
parameters. However, this also presents a key challenge to the
analysis: Our discussion in \S\ref{sec:NH} shows that we cannot assume
that the dust column density is uniform across the FOV traversed by
the echo.  Because image analysis can only measure the scattering
depth $\Delta \tau$, which depends on {\em both} the scattering cross
section {\em and} the column density, it is {\em impossible} to derive
both from an analysis of intensity profiles. In other words: simply by
measuring a slope in the ring intensity as a function of ring- (and
thus scattering-) angle, we cannot distinguish whether this slope is
induced by a change in dust column density with angle {\em or} a slope
in the scattering cross section as a function of scattering angle.

However, for a given cloud, we can expect the photo-electric
absorption column to be proportional to the dust scattering column
\citep[e.g.][]{corrales:15}\footnote{An additional degree of
  uncertainty in the coupling between photo-electric absorption column
  and dust scattering column arises from the uncertainty in the
  dust-to-gas ratio, which may vary by up to a factor of two between
  different locations in the Galaxy \citep{burstein:78}; we do not
  attempt to independently constrain the dust-to-gas ratio in our fits
  because it is degenerate with the unknown fluence of the outburst,
  as discussed in \S\ref{sec:fit_results}.}.  We can then hope to {\em
  spectrally} disentangle the ambiguity between $N_{\rm dust}$ and
$d\sigma/d\Omega$ by spectrally fitting the entire echo, and tying the
scattering and absorption columns of each cloud together by a simple
proportionality constant.  This approach requires an accurate spectral
model of the dust scattering cross section. We will discuss our
spectral analysis of the {\em Swift} echo in the next section.

\section{Spectral analysis of the echo}
\label{sec:spectra}

\subsection{{\tt dscat}: An {\tt XSPEC} Model for Differential Dust
  Scattering Cross Sections}
\label{sec:crosssection}

A spectral fit of the dust scattering echo using equation
(\ref{eq:intensity}) or (\ref{eq:fluence}) requires a spectral model of
the differential scattering cross section as a function of scattering
angle and energy.  No such model exists for the {\tt XSPEC} package
used for fitting the ring spectra in this paper.

We generated table models for $d\sigma/d\Omega$ for the most commonly
used dust distributions and will briefly describe the computational
method used to calculate the cross sections.  Our {\tt dscat}
code\footnote{{\tt dscat} is not yet available for public release,
  given that the ranges in scattering angles and energies covered by
  the table model were tuned to the observations described here; it
  may be made available upon request on a collaborative basis.} for
the differential scattering cross section is based on the
public\footnote{{\tt https://github.com/atomdb/xscat}} {\tt xscat}
package released by \citet{smith:16} that presents {\tt xspec} table
models of {\em integrated} scattering cross sections; that paper
contains a detailed description of the methodology used in calculating
dust scattering cross sections.

Cross sections in this paper are calculated using exact Mie scattering
solutions \citep[rather than using interpolations or employing the
common Rayleigh-Gans approximation, e.g.,][]{mauche:86}, employing the
publicly available\footnote{{\tt ftp://climate1.gsfc.nasa.gov} in
  directory {\tt combe/Single\_Scatt/Homogen\_Sphere/Exact Mie/}} Mie
scattering code {\tt MIEV} \citep{wiscombe:79,wiscombe:80}.  We
include spherical harmonic expansions up to order 250 in {\tt MIEV} to
allow convergence at larger grain sizes.

Scattering cross sections depend on the imaginary part $k$ of the
optical constant $m=n+ik$, which is related to the photo-electric
absorption cross section $\sigma_{\rm ph}$ by the optical theorem
\begin{equation}
  k =n_{\rm gr}\lambda \frac{\sigma_{\rm ph}}{4\pi}
\end{equation}
where $n_{\rm gr}$ is the particle density of the dust grains and
$\lambda$ is the X-ray wavelength of the photons under consideration.

The optical constants $n$ for different grain compositions used as
input to the {\tt MIEV} Mie-scattering code are taken from
\citet{zubko:04}, which themselves are based on the photo-ionization
cross sections provided by \citet{verner:96}, using the Kramers-Kronig
relations to derive $n$.

We computed cross sections in the energy range
$0.5\,{\rm keV} \leq E \leq 5\,{\rm keV}$ in 100 logarithmically
spaced energy bins and over a range of scattering angles between
$500'' \leq \theta_{\rm sc} \leq 4500''$, in 100 logarithmically
spaced angle bins.  The computed model grids are written into {\tt
  FITS} tables and cross sections at specific energies and scattering
angles are calculated in {\tt XSPEC} using logarithmic interpolation
in angle and energy.

Following \citet{smith:16}, we implemented four families of dust
distributions as functions of grain size $a$:
\begin{enumerate}
\item{A generalized Mathis-Rumpl-Nordsieck (1977, MRN77 hereafter)
    distribution, which consists of a combination of silicate and
    graphite grains, each following a distribution with a single slope
    of
    \begin{equation}
      \frac{dN_{\rm MRN}}{da} \propto a^{-\psi}
    \end{equation}
    with a default slope of $\psi_{\rm MRN}=3.5$, extending from
    $a_{\rm min}=0.005\mu{\rm m}$ to a maximum grain size of
    $a_{\rm max}$ with a default value of
    $a_{\rm max,MRN}=0.25\mu{\rm m}$.  The default values correspond
    to the parameter choices of the orignal MRN77 model.

    We calculated cross sections over a range in slope
    $2.5 \leq \psi \leq 5.5$ in 31 linearly spaced bins and over a
    range in maximum particle size
    $0.025\mu{\rm m} \leq a_{\rm max} \leq 8.5\mu{\rm m}$ in 39
    logarithmically spaced bins; within this grid, {\tt dscat}
    interpolates linearly and logarithmically in $\psi$ and $a$,
    respectively.  Cross sections were calculated separately for
    silicate and graphite grains, adding both contributions together
    to calculate the differential cross section per hydrogen atom
    using solar abundances. In spectral fitting, we either fix $\psi$
    and $a_{\rm max}$ across all clouds (thus imposing identical dust
    distributions on all clouds), which we call the MRN1 model, or we
    allow these parameters to vary from cloud to cloud (thus fitting
    eight different distributions), which we call the MRN8 model.}
\item{The 32 separate dust distributions derived and presented
    explicitly in \citet[][WD01]{weingartner:01}, using the parameters
    and model names employed in Tables 1 and 3 of that paper. We refer
    to all of these as WD01 models with names based on the parameters
    of each model. These dust distributions were originally developed
    to describe a range of extinction measurement for Milky Way, LMC,
    and SMC sightlines and span a range in $R_{\rm V}$ values and
    composition.}
\item{The 15 separate dust distributions by \citet[][ZDA04]{zubko:04},
    using the parameters and model names presented in that paper with
    a correction for the $b_{1}$ and $b_{2}$ parameters of the {\tt
      COMP} models.  These dust distributions were originally derived
    to fit Milky Way extinction and IR dust emission constraints and
    contain a range of grain compositions.}
\item{The dust distribution presented in \citet[][WSD01]{witt:01},
    derived to describe the scattering properties of X-ray dust
    scattering halos, using a modified MRN77 distribution with a
    larger maximum grain size of $a_{\rm max,WSD}=2$ and steeper grain
    size distribution of $\psi_{\rm WSD} = 4$ for grains larger than
    $0.4\mu{\rm m}$.}
\end{enumerate}
Each model contains as parameters the scattering angle, the dust
column density, and the photon energy, while the generalized MRN
models also allow $\psi$ and $a_{\rm max}$ to vary.

\subsection{Spectral Modeling of the 2015 V404 Cyg Echo}
\label{sec:spectral_model}

\begin{figure}[t]
  \center\resizebox{\columnwidth}{!}{\includegraphics{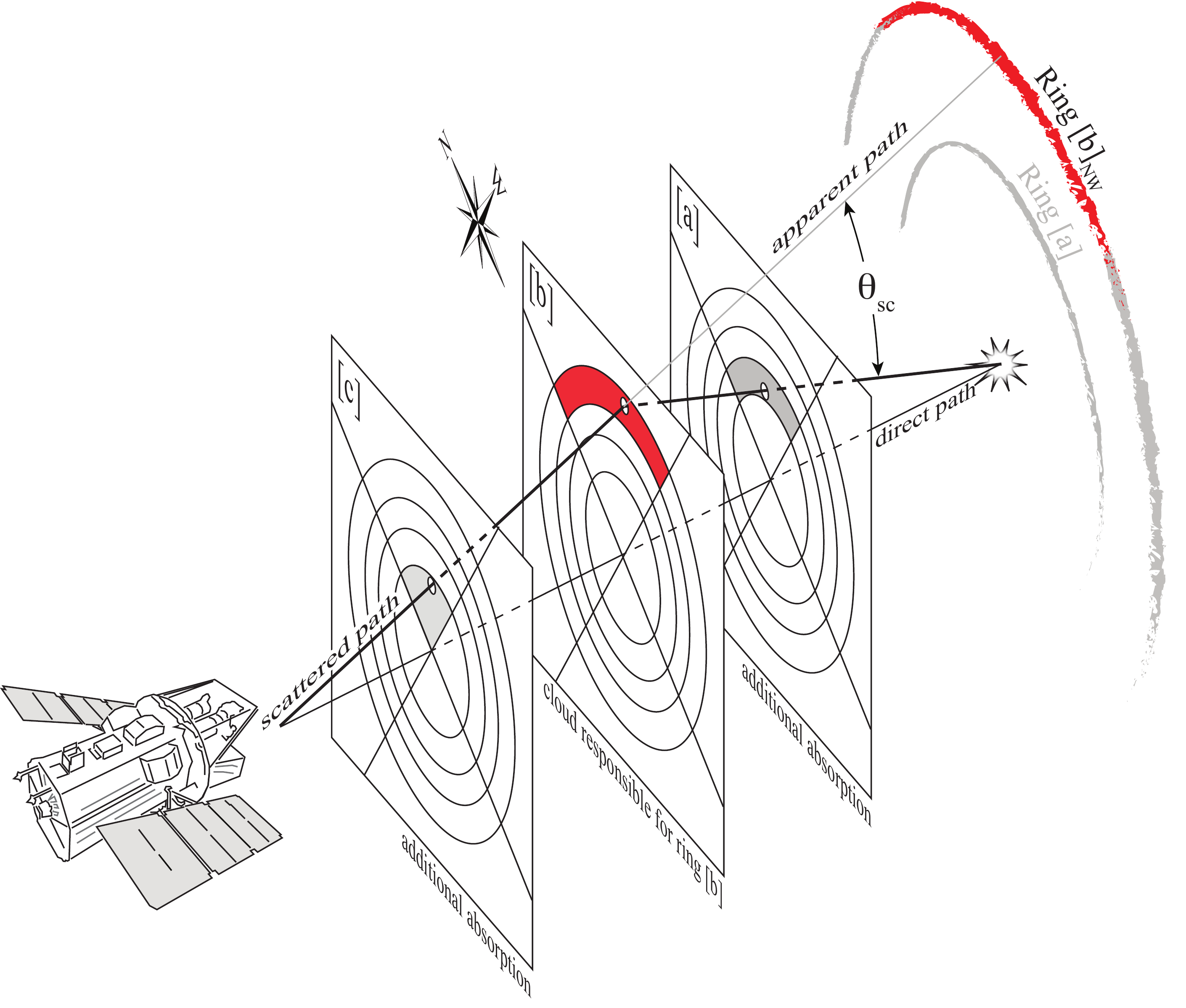}}
  \caption{Cartoon of the generation of X-ray light echoes used in
    constructing the spectral model of the entire echo in
    \S\ref{sec:spectra}. Each distinct cloud along the line of sight
    generates a separate ring (e.g, cloud [b] generates ring [b]), but
    all clouds contribute to the absorption column at different impact
    parameters.  The clouds are divided into sixteen annular sections
    and the association of a ring spectrum with a given section is
    determined from the location where the photon path intersects the
    different clouds. In this simplified example, scattering emission
    from the North-West segment of ring [b] (shown in red) is
    generated in the outermost North-West section of cloud [b], also
    shown in red, while absorption is due to the innermost North-West
    section of cloud [c], the outermost North-West section of cloud
    [b], and the second-innermost North-West section of cloud
    [a].}\label{fig:echo_sketch}
\end{figure}

The scattering geometry of ring [b] (taken as an illustrative example)
is sketched in Fig.~\ref{fig:echo_sketch}. Because clouds cannot be
assumed to be spatially uniform across the field of view, we divide
each cloud into four annuli and each annulus into four quadrants
(North-West, South-West, South-East, and North-East, measured in
equatorial coordinates).  As the light echo expands with time, each
ring (i.e., the echo from each cloud) will sweep across the annuli.
For the sake of computational feasibility, we will assume that the
column density of a cloud is constant {\em within} each of the sixteen
annulus sections.

It is clear from this figure and eq.~(\ref{eq:fluence}) that the
spectrum of (for example) ring $[b]$ in a given ring quadrant
(spanning an azimuthal range of $\Delta \varphi=90^{\circ}=\pi/2)$,
under the assumption of a short flare, is given by
\begin{equation}
  F_{\nu} =e^{-\sigma_{\rm ph}(\nu)\sum_{i={\rm [a]}}^{\rm [h]}N_{\rm
        H,i}(\theta_{i})}\frac{\pi}{2}\frac{c N_{\rm H,[b]}}{x\left(1 -
      x\right)D}\frac{d\sigma_{\rm sc}(\theta_{\rm sc},\nu)}{d\Omega}{\mathcal F}_{\nu}
\label{eq:spectrum}
\end{equation}
where the photo-electric absorption term includes all the gas along
the LOS (including cloud [b]). The same expression holds for all other
clouds.  We assume that most of the cold absorbing gas is located
within the clouds responsible for the echo, but test for the presence
of additional absorbing gas by allowing for an additional amount of
(uniform) foreground absorption of gas and dust not located within
clouds [a]-[h].  Scattering angles are calculated from
eq.~(\ref{eq:theta_scatter}) using cloud positions from Table
\ref{tab:clouds}, assuming the time delay from Table
\ref{tab:obstable}.

We represent the flare spectrum by a simple powerlaw with
normalization and powerlaw index as free parameters.

We incorporate photo-electric absorption using the {\tt PHABS} table
model in {\tt XSPEC} for computational speed.  The column density of
each cloud that is used to calculate the total absorption column in
eq.~\ref{eq:spectrum} will be a function of impact parameter and
azimuthal angle (see \S\ref{sec:NH}).  The impact parameter (in terms
of angle on the sky) can be related to the observed ring angle by
simple geometry using Fig.~\ref{fig:echo_sketch}: for absorption by
clouds closer to the observer than the scattering cloud (for
illustrative purposes, clouds [c] and [b], respectively, in the
cartoon in Fig.~\ref{fig:echo_sketch}), the impact parameter is at the
same on-sky off-axis angle, $\theta_{\rm [c]}=\theta_{\rm [b]}$. For
absorbing clouds further away than the scattering cloud (cloud [a] and
[b] in the cartoon in Fig.~\ref{fig:echo_sketch}, respectively), the
impact parameter is given by the on-sky off-axis angle
$\theta_{\rm [a]} = \theta_{\rm [b]}[x_{\rm [b]}/(1-x_{\rm
  [b]})][(1-x_{\rm [a]})/x_{\rm [a]}]$.

Because, at a fixed time, the scattering emission of all rings
originates from a (convex) ellipsoid (with source and observer at the
two foci), the absorption in fore- and background clouds always occurs
at radii smaller than or equal to the scattering emission for a given
cloud.  E.g., in the example given in Fig.~\ref{fig:echo_sketch} the
dust scattering in cloud [c] that generates ring [c] originates at
larger ring radii than the absorption by cloud [c] of photons
scattered into ring [b] by cloud [b] etc.

We extracted spectra for each {\em Swift} observation\footnote{Because
  of the geometric restrictions of the {\em Chandra} field-of-view, we
  did not jointly fit the spectra of ObsID 17704, since we cannot
  generate region files with comparable coverage to the {\em Swift}
  field-of-view.} and each of the four rings in the four quadrants
(NW, SW, SE, NE).  Given that the echo is soft, we restrict analysis
to the 0.6-5~keV energy range.  Region files were constructed using
ring radii chosen such that all the emission within each region is
dominated by the respective ring, using Fig.~\ref{fig:profiles} to
determine the region of the image dominated by each ring.  We chose
ring boundaries to cover the entire region between the rings, that is,
ring boundaries touch. The boundaries are shown in
Fig.~\ref{fig:stacked}, with the exception that ring [b] extends out
to the inner edge of ring [c] for our spectral extraction.

It is important to note that the dust scattering kernel shown in the
right panel of Fig.~\ref{fig:deconvolution} is significantly broader
than the ring separation during the earlier observations of the echo.
That is, while the bright parts of the ring, which correspond to the
echo of the main flare on MJD 57199.8, are well separated, the echo
from the earlier part of the flare overlaps in part with the main ring
emission at larger angles from clouds closer to the observer.

During the initial echo, this is a relatively small effect, since the
total fluence of the outburst before the main flare is only about 33\%
of the total fluence, and because the strong dependence of ring flux
on scattering angle suppresses the outer emission relative to the main
flare (because the scattering angles for the pre-flare echo are
significantly larger, given that the emission occurred several days
prior to the main flare).  However, we cannot completely neglect this
component, as it represents about a third of the total echo flux
during the later observations, where scattering angles for flare and
pre-flare emission are similar.

For computational feasibility, we account for the pre-flare echo by a
single second spectral component generated by the pre-flare emission
for each ring.  Using the reconstructed soft X-ray lightcurve from
Fig.~\ref{fig:lightcurve}, we calculate an {\em emission-weighted}
pre-flare time of $t_{\rm pre} = $ MJD 57195.3 and calculate
scattering angles for the pre-flare echo from the time delays
referenced to $t_{\rm pre}$.  We add both spectral components to
account for the total dust scattering emission of each cloud.  We fix
the relative fluence of pre-flare and flare emission to be 33\% and
67\% of the total fluence, respectively, corresponding to the
fractions chosen to calculate the emission weighted pre-flare and
flare times.

It is worth emphasizing again that the echo emission from the
different clouds in any given observation extends well beyond the
easily visible rings identified by eye in Fig.~\ref{fig:stacked},
which can be seen from the width of the echo kernel in the right panel
of Fig.~\ref{fig:deconvolution}.  The diffuse flux in the images
therefore contains contributions from the echo on almost all angular
scales outward of the inner edge of ring [a].  Extracting background
spectra from inter-ring regions of a given observation is therefore
{\em inappropriate}, as it would include significant echo
contributions.  Because the relative ring width changes with time
following eq.~(\ref{eq:thinning}), such an approach would not only
underestimate the echo flux, it would also introduce a temporally
varying systematic error.

To avoid systematic biases, background spectra were therefore
extracted from the stacked 2012 blank sky events file for identical
regions used in the spectral extraction, calculating {\tt BACKSCAL}
parameters from the ratio of the vignetted ancillary response files
(ARFs) for observation and background events files, each weighted by
the radial emission profile of the ring.

For the eight clouds identified in the radial profiles, our spectral
model for a given ring section then consists of the following
components in {\tt XSPEC}:
\begin{eqnarray} 
\lefteqn{{\tt phabs}_{\rm foreground}\times\left(\prod_{k={\rm
        [a]}}^{\rm [h]}{\tt phabs}_{k}\right)}\nonumber \\
  & &  \ \ \ \ \ \ \ \times \left({\tt
      const}_{\rm flare}\times {\tt dscat}_{\rm flare} + {\tt
      const}_{\rm pre}\times{\tt dscat}_{\rm
      pre}\right){\color{black}} \nonumber \\
  & & \ \ \ \ \ \ \ \times \ {\tt powerlaw} \nonumber
  \label{eq:xspec_model}
\end{eqnarray}
where each of the ${\tt phabs}$ components in the bracket correspond
to one of the eight clouds and the additional component allows for
uniform additional foreground absorption, tied across all spectra,.
The {\tt dscat} models represent one of the four dust implementations
(MRN77, WD01, ZDA04, WSD04) listed above, evaluated at the (fixed)
scattering angles of the main flare and the mean pre-flare value,
respectively.

The constants contain the covering fraction of each ring (calculated
using the exposure map of the observation and the region file of the
ring section), the ratio of pre-flare to flare fluence, and the
remaining terms from eq.~(\ref{eq:spectrum}), $(\pi/2) c/[x(1-x)D]$.

Note that both the photo-electric absorption optical depth
$\tau_{\rm phabs}$ and the dust scattering optical depth
$\tau_{\rm sc}$ depend on the Hydrogen column density. The ratio of
$\tau_{\rm phabs}$ to $\tau_{\rm sc}$ depends on the
dust-model-dependent dust-to-gas ratio. Our analysis is unable to
constrain the dust-to-gas ratio independently, as it also correlates
with the normalization constant of the powerlaw (which cannot be
separately determined).

We fit all ring spectra {\em simultaneously}, tying all powerlaw
normalizations and slopes together, and tying the column density in
each of the sixteen cloud sections for each cloud together across all
spectra.  From the 50 {\em Swift} observations, each containing
sixteen segments for each of the eight clouds, we include a total of
857 spectra that satisfy our 40 count minimum in the fit. The median
number of counts per spectrum is 82, while the mean number of counts
is 120.  Spectra were grouped to a minimum of 20 counts per bin.  Not
every cloud segment is covered by a ring spectrum of sufficient
counts; the outermost sections of rings [f], [g], and [h] are only
represented in some of the quadrants.

The resulting fits involve a total of about 21,000 parameters
(depending on the exact model used), almost all of which are tied or
fixed. For example, our fiducial MRN1 model with a single (free)
particle slope $\psi$ and a single free maximum particle size
$a_{\rm max}$ has total of {\color{black}{127}} {\em free} parameters,
{\color{black}{122}} of which are the dust column densities for the
different cloud segments (and thus directly proportional to the
scattered flux), compared to a total of 4897 spectral bins used for
fitting for a total of 4770 degrees of freedom.

Three-sigma uncertainties in fit parameters were computed using the
{\tt error} or {\tt steppar} commands in {\tt XSPEC}.

\section{Discussion}
\label{sec:discussion}

\subsection{Spectral Constraints on Dust Models and Cloud Columns}
\label{sec:fit_results}

The fit statistics for all spectral models employed in our fits are
listed in Table \ref{tab:dustfits}.  The table also lists the
approximate slope $\alpha$ of the scattering cross section with
scattering angle at 1.5 keV (characteristic of the median photon
energy).

The fit statistic (reduced chi-square\footnote{We tested the
  sensitivity of our results to the specific fit statistic used in the
  case of the MRN1 and MRN8 model fits and found that fit parameters
  derived from both chi-square and Cash \citep[preferred for spectral
  fits with low counts, e.g., ][]{nousek:89} statistic are consistent
  within the stated uncertainties.}) varies from acceptable (1.05) to
clearly rejected (1.7) across the models we fit to the data.  A
standard MRN77 model produces a marginally acceptable fit, with
reduced chi-square of 1.09.  The best fit WD01 model (reduced
chi-square of 1.08) turns out to be the SMC bar dust model with an
$R_{\rm V}$ of 2.9, while the bestfit ZDA04 models are the bare grain
Graphite-Silicate models (reduced chi-square of 1.07-1.08).

We find that our fits do not require an additional foreground
absorption component, placing a three-sigma upper limit of
$N_{\rm foreground} < 6\times 10^{20}\,{\rm cm^{-2}}$ on the amount of
foreground absorption not related to the dust-scattering clouds.

Generally, models with steeper slopes in the dust distribution and
correspondingly shallower slopes of the cross section with scattering
angle, i.e., smaller $\alpha$, provide better fits, as can be seen
from the table.

We can quantify the preference for steeper dust distributions by
fitting generalized MRN models with both slopes and maximum grain size
left as free parameters.  The MRN1 model fitting a single size
distribution to all eight clouds listed in Table \ref{tab:dustfits} is
our fiducial model. We also include a model that allows the maximum
grain size and slope of the dust distribution to vary from cloud to
cloud, called MRN8 in Table \ref{tab:dustfits}.  

We prefer the fiducial MRN1 model because cloud parameters are
spectrally coupled especially for rings [a] and [b]. From
Fig.~\ref{fig:deconvolution}, we can see that 80\% of the ring flux is
contained within a ring of width $\Delta \theta/\theta \sim $20\%,
while the rings are typically separated by about 12\% to 15\% in
radius only (with the exception of rings[b] and [c], which are
separated by 50\% and therefore do not overlap, which we made use of
in \S\ref{sec:lightcurve}).  Therefore, during the earlier
observations, the ring fluxes are contaminated by about 10\% of the
flux from the next-innermost ring. This contamination will affect the
temporal/angular evolution of the two rings that are coupled in
opposite ways (i.e., ring [a] is expected to bleed into the spectral
extraction region of ring [b], thus lowering the flux of ring [a] and
increasing the flux of ring [b]).  By tying all dust models together,
we can expect that the effect cancels to lowest order.

For our MRN1 model (tying all the dust slopes and $a_{\rm max}$
together across the clouds, thus fitting identical grain size
distributions to all clouds), we find a best-fit slope of
\begin{equation}
  \psi_{\rm MRN1} =\frac{d\ln{(dN)}}{d\ln{a}} = 3.95^{+0.06}_{-0.18}
\end{equation}
while the maximum grain size is poorly constrained, with a best fit
value of $a_{\rm max} \sim 4.0$.  We can place a three-sigma lower
limit of
\begin{equation}
  a_{\rm max,MRN1} > 0.15\mu{\rm m}
\end{equation}
where $\psi$ and $a_{\rm max}$ are somewhat degenerate, i.e., steeper
grain distributions allow larger maximum grain sizes
\citep{corrales:13}.

\begin{figure}[t]
  \center\resizebox{0.8\columnwidth}{!}{\includegraphics{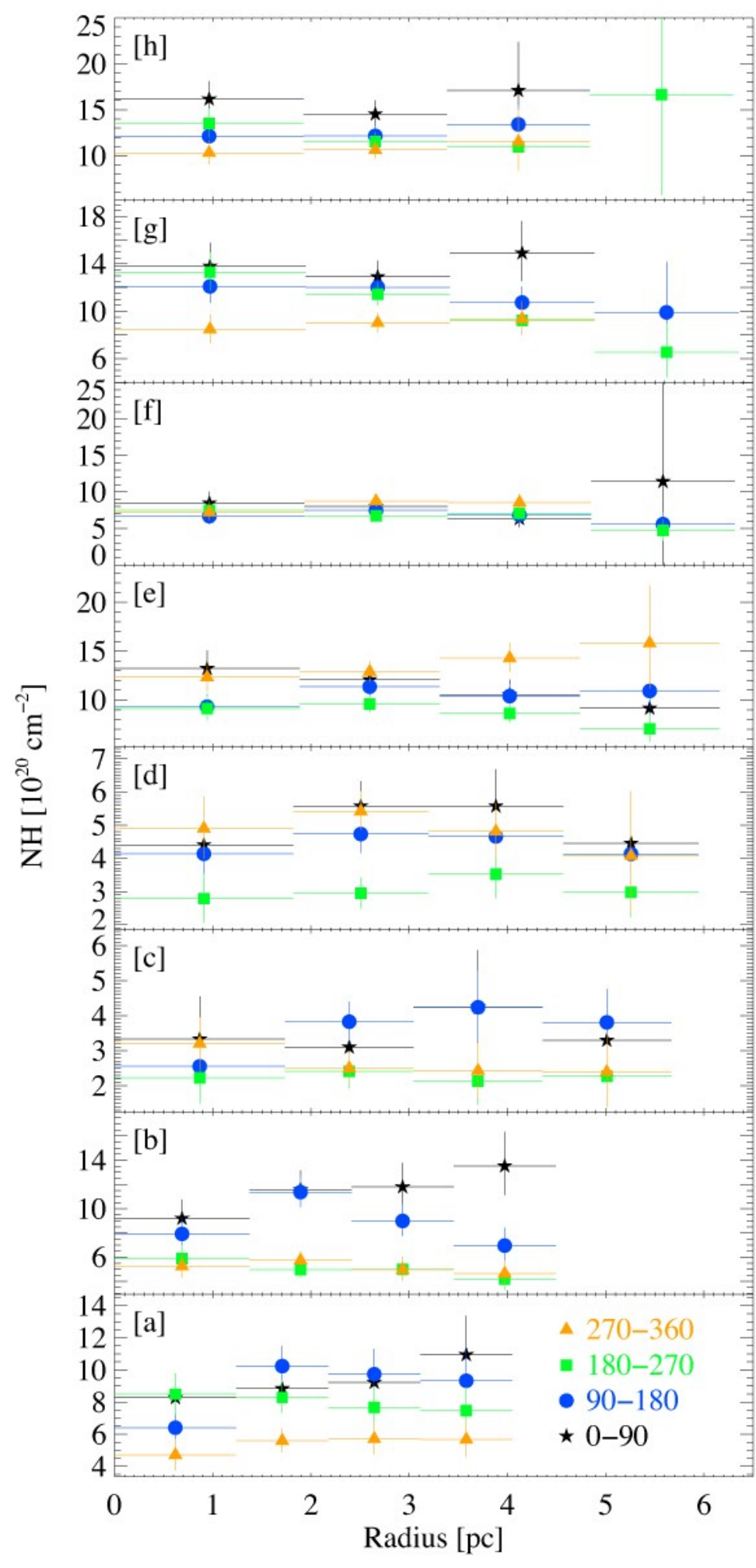}}
  \caption{Column densities of clouds [a]-[h] in the four equatorial
    quadrants determined from spectral fitting of our fiducial MRN1
    model described in \S\ref{sec:fit_results}.  Position angles are
    measured clockwise relative to North, with black stars, blue dots,
    green squares, and orange triangles corresponding to the
    North-West, South-West, South-East, and North-East quadrants,
    respectively.}\label{fig:cloud_nh}
\end{figure}

From the fits, we can also derive the dust column density in the
sixteen ring sections we employed for each of the clouds.  For our
fiducial MRN1 model, the best-fitting spatial dust distribution for
each cloud is shown in Fig.~\ref{fig:cloud_nh}, which plots
$N_{\rm H}$ as a function of angle and impact parameter from the
line-of-sight in physical distance.  Note that, while the angular
sizes of the clouds on the sky are very different, the physical cloud
sizes covered by the echo are comparable.  The figure confirms the
discussion in \S\ref{sec:NH}: Many of the clouds have higher
$N_{\rm H}$ in the direction toward the Galactic plane (in quadrant
0-90), especially cloud [b], which shows a roughly fourfold increase
in the NW quadrant at large ring radii.  Mean cloud columns and
variances derived from the fit are listed in Table \ref{tab:clouds}.

Given the clear column density peak of cloud [b] in the North-West
quadrant identified both spectrally and in
Fig.~\ref{fig:cloud_images}, we investigated the Pan-STARRS extinction
data in that direction and found a very large excess in extinction at
the location of the peak in $N_{\rm H}$: The differential extinction
at the location of column density peak near
{\color{black}{(RA=20:23:41,DEC=33:56:39)}} in the distance range of
cloud [b] from 2.0-2.5 kpc is $\Delta E(B-V)=0.95$, while the typical
value in the South-Eastern direction (i.e., near the equi-distant
location in the opposite direction from V404 Cyg) is around
$\Delta E(B-V) \sim 0.2$, confirming that the column density peak
detected in the echo and absorption data is real and reflected in the
visible extinction.  A smoothed contour of the Pan-STARRS extinction
map in the 2.0-2.5 keV distance range is plotted as a dashed contour
in Fig.~\ref{fig:cloud_images}b, showing clear spatial coincidence
with the excess in dust column derived from the echo.

It is clear from Figs.~\ref{fig:angular_profiles} and
\ref{fig:cloud_nh} that the assumption of uniform dust column for a
given cloud is not justified and results derived under the assumption
of uniform column, especially regarding the slopes of the scattering
cross section with angle (which relate to the slope of the grain size
distribution), may be unreliable. We can test the assumption of
uniform column per cloud by fitting the same spectra as above, but
tying the columns of each cloud segment together.  This reduces the
fit to twelve free parameters, eight of which are the cloud column
densities. This model is statistically rejected, with a reduced
chi-square of 1.6.  The inferred parameters of the particle
distribution for this fit are closer to the standard MRN77 model, with
a slope of $\psi=3.76$ and $a_{\rm max} = 0.40$, corresponding to
$\alpha_{\rm 1.5keV} \sim 3.2$.

Using our MRN8 model, we test for evidence of variations in dust
properties from cloud to cloud. The fits statistic is marginally
better (1.046) than in the case of the {\color{black}{uniform dust}}
fit of model MRN1 (1.063).  Slopes and maximum grain sizes determined
from the fits to the MRN8 model are listed in Table \ref{tab:clouds}.

{\color{black}{We find grain size distributions with power-law index
    significantly steeper than 3.5 in clouds [a], [b], and [h], and a
    marginal preference for steeper distributions for clouds [c], [e],
    and [h]}}. Thus, while our fits do not support the suggestion of a
clear gradient in cloud properties toward V404 Cyg as suggested by
\citet{vasilopoulos:16}, we confirm their finding that the dust
properties of clouds [a] and [b] are best fit by steeper dust
distributions than a typical MRN model.

Rings [a] and [b] sample the largest scattering angles (the range
$2500 < \theta_{\rm sc} < 3400$ is explored only {\color{black}{by}}
rings [a] and [b]) and thus the smallest grains.  Because they are
brightest (due to the $(1-x)^{-2}$ brightening factor), rings [a] and
[b] may dominate the global fits of our MRN1 model.  For both reasons,
we cannot test whether the steeper slope preferred for clouds [a] and
[b] is a result of a different dust composition in clouds [a] and [b]
or whether all clouds require an excess of small grains.

To test whether the properties of the dust correlate with column
density, we allowed the slope and cutoff in the North-West quadrant of
rings [a] and [b] to vary independently in a variant of our MRN8
model.  We find that the best-fit slopes in the North-West quadrant
are even steeper (with slopes of $\psi \sim 4.3$ and $\psi \sim 4.0$
for rings [a] and [b], respectively).  However, given the
well-localized extent of ring [b], which is even smaller than the
annular extraction regions used in the fits, it is possible that this
steepening is a residual artifact of the varying column as a function
of distance.

Given that a standard MRN77 model can describe the data reasonably
well, with a marginally higher reduced chi-square (1.09 compared to
1.05), we feel that strong claims about a slope $\psi$ and cutoff
$a_{\rm max}$ that differ significantly from a standard MRN model have
to be taken with a grain of salt.  We can, however, rule out models
with, on average, significantly shallower dust slopes (i.e., larger
$\alpha$), in particular, some of the $R_{\rm V} > 3.1$ models from
WD01 and some of the composite ZDA04 models, given their poor fit.

A generally robust conclusion from our fits is that the data are
explained sufficiently well by simple dust distributions of bare
silicate and graphite grains. The more complex composite ZDA models
and models of covered grains generally do not provide acceptable fits.

The average spectrum of the outburst is derived from the powerlaw
component in the model. For our fiducial MRN1 model, we find a
spectral slope of
\begin{equation}
  \Gamma_{\rm V404} = 2.2^{+0.15}_{-0.15}
  \label{eq:gamma}
\end{equation}
and a total (un-absorbed) fluence [using eq.~(\ref{eq:fluence})] of
\begin{equation}
  {\mathcal F}_{1-3keV} \sim 0.02\,{\rm ergs\,cm^{-2}}
  \label{eq:404_fluence}
\end{equation}
while the MRN8 model gives a spectral slope of
$\Gamma =2.1^{+0.15}_{-0.15}$ and a fluence of
${\mathcal F} \sim 0.015\,{\rm ergs\,sm^{-2}}$.

The fluence determined from the dust fits is larger than, but
comparable to, the inferred {\em INTEGRAL} fluence in the 1-3 keV band
from Fig.~\ref{fig:lightcurve} [using the fitted spectral slope from
eq.~(\ref{eq:gamma}) to extrapolate the 3-7 keV fluence to the 1-3 keV
band] of $ {\mathcal F}_{JEMX,1-3keV} \sim 0.012\,{\rm ergs\,cm^{-2}}$
\citep{kuulkers:15}.  We note, however, that the normalization of the
powerlaw in eq.~(\ref{eq:404_fluence}) depends on the dust-to-gas
ratio, which we cannot constrain independently from the powerlaw
normalization of the flare.  Also, the spectral slope is likely not
constant with energy, and thus an extrapolation from the {\em
  INTEGRAL} band to the 1-3 keV band may be inaccurate. Additionally,
the gaps in the {\em INTEGRAL} lightcurve suggest that the hard X-ray
{\em INTEGRAL} fluence itself is likely an underestimate. Thus, we do
not quote uncertainties in ${\mathcal F}$, given that they are likely
dominated by the systematics.

\subsection{Comparison with the 2014 Circinus X-1 Light Echo}
It is worth briefly comparing the 2015 V404 Cyg light echo to the 2014
echo seen around Circinus X-1, reported in \citet{heinz:15}.

The fluence of the June 2015 outburst was about a factor of two lower
than the fluence of the 2014 Circinus X-1 outburst. Given the larger
column density toward Circinus X-1, the light echo reported in
\citet{heinz:15} was therefore intrinsically brighter compared to the
V404 Cyg echo. Because of the larger distance by a factor of four and
the correspondingly smaller angular size of the echo on the sky, the
echo was observable for longer delay times (60 days after the outburst
in XMM ObsID 0729560501 compared to the last unambiguous detection of
the V404 Cyg echo 40 days after the end of the outburst in ObsID
00031403115).

The V404 Cyg outburst was significantly shorter and dominated by a
single (possibly double-peaked) strong flare at the end of the
outburst, while the Cir X-1 outburst consisted of longer persistently
bright emission, with an interspersed quiescent binary orbit.  As a
result, the rings observed in the V404 Cyg echo are much narrower than in
the Cir X-1 echo.

Clearly, the well-known distance to V404 Cyg makes it a more accurate
probe of the interstellar dust distribution. The more complete
sampling of the temporal evolution of the echo allows more accurate
constraints on dust scattering cross sections and thus dust properties
to be placed.

Unfortunately, the dust sampled toward V404 Cyg by the echo lies
almost exactly along the tangent of the Galactic rotation curve, and CO
and HI observations of the source provide very little kinematic
leverage, unlike in the case of Circinus X-1.  Still, high-resolution
imaging of the region to look for CO peaks at the locations of the
clouds identified here and in \citet{vasilopoulos:16} could provide
additional diagnostics for cloud and dust properties, especially in
the case of the well-defined peak of the cloud [b]-complex.

\section{Conclusions}
\label{sec:conclusions}
We presented a combined analysis of the {\em Chandra} and {\em Swift}
observations of the 2015 X-ray light echo of V404 Cyg.  By devising a
new stacking technique for light echo data, we were able to generate a
deep image of the echo from the combined {\em Swift} images that
reveals eight distinct echo rings, corresponding to eight separate
dust clouds along the line of sight.

Analysis of the innermost rings in the {\em Chandra} observations
shows that the dust column densities in the corresponding clouds are
non-uniform across the field of view, invalidating the default
assumption of uniform column density for a given cloud in the analysis
of dust scattering echoes.
 
Cross correlations of the radial intensity profiles of the echo place
the majority of the fluence from the June 2015 outburst of V404 Cyg in
a single major flare on MJD 51799.8.  Using the azimuthal variation in
intensity of ring [b], we reconstructed the soft X-ray lightcurve of
the outburst, showing that the main flare contained approximately two
thirds of the outburst fluence.

We reconstructed the dust distribution toward V404 Cyg by
deconvolving each of the radial intensity profiles with the
reconstructed outburst lightcurve, following the technique developed
in \citet{heinz:15}, deriving locations for each of the eight dust
clouds from a weighted average of all the {\em Swift} deconvolutions,
which are consistent with the locations derived from {\em Chandra}
ObsID 17704.

We showed that the non-uniformity in cloud dust column across the
field of view presents a significant challenge for fitting models of
dust scattering to the dust scattering intensity as a function of
time/scattering angle.

In order to mitigate this difficulty and to self-consistently fit echo
spectra, we developed a new {\tt XSPEC} model of the differential dust
scattering cross section for four commonly used dust distributions
from the literature commonly used to fit cross sections: A generalized
MRN77 model with varying slope and maximum grain size, 32 different
WD01 models and 15 different ZDA04 models, as well as a single WSD04
model.  Given the clear variation of dust column across the field of
view, we allowed the dust column to vary with radius and angle in our
spectral fits.

We presented joint fits of all {\em Swift} spectra that make use of
the fact that photo-electric absorption and dust scattering column are
proportional to allow us to vary the column across cloud radius and
azimuthal angle in each cloud, employing sixteen annular sections for
each cloud. We showed that a standard MRN77 model (with a slope of
3.5) provides a reasonable global fit (reduced chi-square of 1.09),
and that fits prefer somewhat steeper dust distributions, with slopes
closer to 4.0, and larger grain size cutoffs; in fact, we can only
place a lower limit of $a_{\rm max} > 0.15\mu{\rm m}$ on the upper
cutoff.

 By allowing dust parameters to vary from cloud to cloud in our fits,
we find no evidence for systematic trends in grain properties as a
function of distance from V404 Cyg. However, our spectral fits confirm
the suggestion by \citep{vasilopoulos:16} that the two clouds closest
to V404 Cyg (clouds [a] and [b]) are best fit by steeper dust
distributions.  We find that the region of highest column density in
those clouds in the direction toward the Galactic plane is best fit by
even steeper dust distributions, with slopes between 4.0 and 4.3,
suggesting an excess of small grains inside the dense core of those
clouds.

More generally, we find that simple dust compositions consisting of
silicate and graphite grains provide acceptable fits, while more
complex models, and models with very shallow dust distributions are
statistically ruled out, as are models with uniform column density
across the FOV.

The observations show that a combination of deep, high-resolution {\em
  Chandra} images (at least two separate epochs spaced sufficiently
far apart in time to allow accurate cross-correlation of the radial
profiles) and early and frequent {\em Swift} pointings provide a
powerful tool in the study of dust scattering echoes. Our analysis
demonstrates that even short {\em Swift} observations, if sufficiently
frequent, can be stacked and analyzed jointly; thus, total observing
time (especially at late times during the echo) is more important than
length of individual segments, making best use of the scheduling
flexibility of {\em Swift}.

Future campaigns to observe scattering echoes will be most successful
when employing these resources strategically.  In particular, by
placing the point source in the chip gap between ACIS S and ACIS I,
the {\em Chandra} field-of-view can be maximized while mitigating risk
to the instruments from re-brightening of the source. HETG
observations of faint diffuse echoes are severely hampered by the lack
of available blank sky backgrounds and low sensitivity in the crucial
1-2 keV band and cannot be substituted for dedicated imaging
observations.

\acknowledgements{We would like to thank Bob Benjamin and Erik
  Kuulkers for helpful discussions.  We are grateful to CXO Director
  Belinda Wilkes for granting two Director's Discretionary Time
  observations of the echo, as well as the {\em Chandra} observations
  and scheduling team for their rapid turnaround and support in
  devising successful and safe observing parameters. This research has
  made use of data obtained from the Chandra Data Archive and the {\em
    Chandra} Source Catalog, and software provided by the Chandra
  X-ray Center (CXC) in the application packages {\tt CIAO}, {\tt
    ChIPS}, and {\tt Sherpa}.  SH acknowledges support for this work
  was provided by the National Aeronautics and Space Administration
  through Chandra Award Number GO4-15049X and RS acknowledges support
  through Chandra Award Number TM4-15002X, issued by the {\em Chandra}
  X-ray Observatory Center, which is operated by the Smithsonian
  Astrophysical Observatory for and on behalf of the National
  Aeronautics Space Administration under contract NAS8-03060. We
  acknowledge the use of public data from the Swift data archive.}

{\footnotesize{\begin{deluxetable}{lllclc}
\tablecaption{Observation parameters\label{tab:obstable}}
 
\tablehead{
\colhead{ObsID} & 
\colhead{Telescope} & 
\colhead{Mode} & 
\colhead{Obs Date [MJD]} & 
\colhead{$\Delta t [d]$} & 
\colhead{Exp [ksec]}}

\startdata
00031403071 & Swift & XRT-PC & 57203.46 & 3.68 & 1.0\\
00033861006 & Swift & XRT-PC & 57205.46 & 5.69 & 1.7\\
00031403072 & Swift & XRT-PC & 57205.80 & 6.03 & 0.9\\
00031403074 & Swift & XRT-PC & 57206.67 & 6.89 & 0.8\\
00033861007 & Swift & XRT-PC & 57207.39 & 7.61 & 1.5\\
00031403076 & Swift & XRT-PC & 57207.53 & 7.75 & 1.2\\
00033861008 & Swift & XRT-PC & 57208.38 & 8.61 & 1.5\\
00031403079 & Swift & XRT-PC & 57208.93 & 9.15 & 1.5\\
00031403078 & Swift & XRT-PC & 57208.99 & 9.22 & 0.9\\
00031403080 & Swift & XRT-PC & 57209.52 & 9.75 & 0.9\\
00081751001 & Swift & XRT-PC & 57209.72 & 9.95 & 1.8\\
00031403083 & Swift & XRT-PC & 57210.31 & 10.54 & 0.9\\
00031403081 & Swift & XRT-PC & 57210.38 & 10.61 & 0.9\\
00031403084 & Swift & XRT-PC & 57211.51 & 11.74 & 0.9\\
00031403085 & Swift & XRT-PC & 57211.79 & 12.01 & 1.7\\
00031403086 & Swift & XRT-PC & 57212.79 & 13.02 & 2.0\\
00031403087 & Swift & XRT-PC & 57213.05 & 13.28 & 0.7\\
00031403088 & Swift & XRT-PC & 57213.12 & 13.35 & 0.5\\
00031403089 & Swift & XRT-PC & 57213.19 & 13.42 & 0.2\\
00031403090 & Swift & XRT-PC & 57213.31 & 13.54 & 0.4\\
00031403091 & Swift & XRT-PC & 57213.38 & 13.60 & 0.7\\
00031403092 & Swift & XRT-PC & 57213.44 & 13.67 & 0.5\\
00031403093 & Swift & XRT-PC & 57213.52 & 13.74 & 1.2\\
00031403094 & Swift & XRT-PC & 57213.65 & 13.87 & 1.5\\
00031403095 & Swift & XRT-PC & 57213.71 & 13.94 & 1.6\\
00031403096 & Swift & XRT-PC & 57213.77 & 14.00 & 0.7\\
00033861009 & Swift & XRT-PC & 57214.33 & 14.56 & 1.5\\
00031403097 & Swift & XRT-PC & 57214.60 & 14.83 & 1.9\\
17701 & Chandra & ACIS-S/HETG & 57214.79 & 15.01 & 39.5\\
00031403098 & Swift & XRT-PC & 57215.65 & 15.87 & 0.9\\
00031403099 & Swift & XRT-PC & 57216.18 & 16.41 & 0.5\\
00031403100 & Swift & XRT-PC & 57217.83 & 18.06 & 0.7\\
00033861011 & Swift & XRT-PC & 57218.08 & 18.30 & 2.0\\
00033861010 & Swift & XRT-PC & 57218.51 & 18.74 & 5.3\\
00031403101 & Swift & XRT-PC & 57219.34 & 19.57 & 2.0\\
00031403102 & Swift & XRT-PC & 57220.97 & 21.19 & 0.8\\
00031403103 & Swift & XRT-PC & 57221.34 & 21.56 & 0.9\\
00031403104 & Swift & XRT-PC & 57222.24 & 22.46 & 1.7\\
00031403105 & Swift & XRT-PC & 57223.26 & 23.48 & 1.7\\
00031403107 & Swift & XRT-PC & 57223.53 & 23.75 & 9.2\\
00031403109 & Swift & XRT-PC & 57226.41 & 26.64 & 1.5\\
00031403108 & Swift & XRT-PC & 57226.65 & 26.87 & 6.8\\
17704 & Chandra & ACIS-S/I & 57228.95 & 29.18 & 28.4\\
00031403111 & Swift & XRT-PC & 57231.24 & 31.47 & 11.2\\
00031403112 & Swift & XRT-PC & 57231.53 & 31.76 & 0.5\\
00031403113 & Swift & XRT-PC & 57235.52 & 35.75 & 10.2\\
00031403115 & Swift & XRT-PC & 57239.35 & 39.57 & 9.7\\
00031403116 & Swift & XRT-PC & 57243.57 & 43.80 & 1.4\\
00031403117 & Swift & XRT-PC & 57247.12 & 47.34 & 1.5\\
00031403118 & Swift & XRT-PC & 57251.04 & 51.27 & 1.2\\
00031403119 & Swift & XRT-PC & 57255.21 & 55.43 & 1.3\\
00031403120 & Swift & XRT-PC & 57259.46 & 59.69 & 0.3\\
\enddata
\end{deluxetable}
}}

\begin{deluxetable}{llll}
  \tablecaption{{\em Chandra} PSF fits to rings [a]-[d] \label{tab:psf}}
  
  \tablehead{
    \colhead{Ring} & 
    \colhead{centroid [arcmin]} & 
    \colhead{distance [kpc]} &
    \colhead{line-of-sight FWHM [pc]} 
  }
  
\startdata
${\rm [a]}$ & $5.43\pm 0.05$ & $2.13\pm 0.01$ & $<7.5$ \\
${\rm [b]}$ & $6.46\pm 0.03$ & $2.03\pm 0.01$ & $<8.8$ \\
${\rm [c]}$ & $9.48\pm 0.02$ & $1.74\pm 0.01$ & $<12.6$ \\
${\rm [d]}$ & $10.60\pm 0.02$ & $1.63\pm 0.01$ & $<24.6$  \\
\enddata
\end{deluxetable}

\newpage
\clearpage

\begin{deluxetable}{lllllllll}
  \tablecaption{Ring [a] and [b] fits from ObsID 17704\label{tab:17704_table}}
  
  \tablehead{    \colhead{Ring} & 
    \colhead{$N_{\rm H,NW}$} &
    \colhead{$N_{\rm H,SW}$} &
    \colhead{$N_{\rm H,SE}$} &
    \colhead{$N_{\rm H,NE}$} &
    \colhead{$\phi_{\rm NW}$} &
    \colhead{$\phi_{\rm SW}$} &
    \colhead{$\phi_{\rm SE}$} &
    \colhead{$\phi_{\rm NE}$}
  }
  \startdata
  ${\rm [a]}$  &
  $1.44^{+0.34}_{-0.28}$  & 
  $1.16^{+0.23}_{-0.20}$ & 
  $1.02^{+0.18}_{-0.15}$ & 
  $0.98^{+0.30}_{-0.24}$ & 
  $1.60^{0.62}_{0.46}$ &
  $2.09^{0.53}_{0.44}$ &
  $1.46^{0.31}_{0.26}$ &
  $0.69^{0.30}_{0.23}$   \\
  ${\rm [b]}$  &  
  $1.49^{+0.26}_{-0.23}$ & 
  $1.02^{+0.32}_{-0.26}$ & 
  $0.97^{+0.20}_{-0.17}$ & 
  $1.05^{+0.27}_{-0.22}$ & 
  $4.28^{+0.98}_{-0.81}$ &
  $1.05^{+0.58}_{-0.40}$ &
  $1.46^{+0.30}_{-0.24}$ &
  $1.14^{+0.42}_{-0.32}$  \\
  \enddata

\end{deluxetable}

\begin{deluxetable}{llccccc}
\tablecaption{Cloud properties\label{tab:clouds}}
 
\tablehead{
\colhead{Cloud} & 
\colhead{Distance\tablenotemark{a} $\left[kpc\,\frac{D_{V404}}{2.39\,{\rm kpc}}\right]$} & 
\colhead{Depth [pc]} &  
\colhead{$N_{\rm H}\tablenotemark{b}\ [10^{20}\,{\rm cm^{-2}}$]} & 
\colhead{$\psi_{\rm MRN8}$\tablenotemark{c}} &
\colhead{$a_{\rm max,MRN8}$}}

%
%
%
%
%
%
%

\startdata
${\rm [a.1]}$ & $ 2.126\pm0.012_{\rm CXO} $ & $ 11.48\pm1.24 $ & $7.9 \pm 1.8 $ & $4.08^{+0.19}_{-0.19}$ & $>0.19 (\sim4.6) $ \\
${\rm [a.2]}$ & $ 2.088\pm0.009_{\rm CXO} $ & $ 9.13\pm2.51 $ & $--- $ & --- & --- \\
${\rm [b.1]}$ & $ 2.053\pm0.004_{\rm CXO} $ & $ 3.70\pm1.08 $ & $--- $ & --- & --- \\
${\rm [b.2]}$ & $ 2.039\pm0.011_{\rm CXO} $ & $ 11.12\pm3.74 $ & $7.6 \pm 3.0 $ & $3.90^{+0.20}_{-0.24}$ & $>0.16 (\sim 2.9)$\\
${\rm [c]}$ & $ 1.745\pm0.004_{\rm CXO} $ & $ 3.93\pm0.81 $ & $ 3.0\pm0.7 $ & $3.68^{+0.40}_{-0.74} $ & $>0.11 (\sim 2.2)$ \\
${\rm [d]}$ & $ 1.633\pm0.006_{\rm CXO} $ & $ 5.62\pm0.55 $ & $ 4.3\pm 0.9 $ & $3.53^{+0.41}_{-0.43}$ & $0.14^{+8.5}_{0.03} $ \\
${\rm [e]}$ & $ 1.490\pm0.021 $ & $ 21.95\pm0.57 $ & $ 11.0\pm 2.3 $ & $3.67^{+0.21}_{-0.21} $ & $0.15^{+0.08}_{-0.01} $ \\
${\rm [f]}$ & $ 1.334\pm0.037 $ & $ 37.30\pm2.82 $ & $ 7.4 \pm1.6 $ & $ 3.57^{+0.34}_{-0.23} $ & $0.20^{+1.70}_{-0.05}$ \\
${\rm [g]}$ & $ 1.169\pm0.022 $ & $ 21.95\pm0.57 $ & $ 11.0\pm 2.4 $ & $3.61^{+0.22}_{-0.20}$ & $0.18^{+0.21}_{-0.02} $ \\
${\rm [h]}$ & $ 1.031\pm0.057 $ & $ 57.00\pm3.54 $ & $ 13.1\pm 2.3 $ & $3.92^{+0.16}_{-0.16} $ & $>0.23 (\sim 8.14)$ \\
\enddata
\tablenotetext{a}{Distances and widths for clouds $[a]-[d]$ are
  measured from {\em Chandra} ObsID 17704, as denoted by a CXO
  subscript, distances and widths for clouds [e]-[h] are measured from
  the deconvolution of the {\em Swift}; both are shown in
  Fig.~\ref{fig:column}. Statistical distance uncertainties are
  dominated by the width of the component listed in column 3 and do
  not include the 6\% uncertainty in the distance to V404 Cyg.}
\tablenotetext{b}{Column densities and uncertainties are the mean
  values and 3-sigma uncertainties, respectively, from our fiducial
  MRN1 fit of a single generalized MRN model with uniform dust
  parameters from cloud-to-cloud, shown in Fig.~\ref{fig:cloud_nh}.}
\tablenotetext{c}{Dust slope and upper size cutoff are the best-fit
  values and associated 3-sigma uncertainties from the MRN8 fit of a
  generalized MRN distribution allowing dust parameters to vary from
  cloud-to-cloud.}
\end{deluxetable}

\newpage
\clearpage

\begin{deluxetable}{llcccc}
\tablecaption{Dust model fits\label{tab:dustfits}}
 
\tablehead{
\colhead{Model name} & 
\colhead{Reference} & 
\colhead{$\chi^2$} & 
\colhead{${\rm D.O.F.}$} & 
\colhead{$\chi^{2}_{\rm red}$} & 
\colhead{$\alpha_{1.5\,{\rm keV}}$\tablenotemark{a}}}

\startdata
MRN & MRN77 & 5180.34 & 4772 & 1.09 & 3.29\\
MRN1 & MRN77 & 5071.01 & 4770 & 1.06 & 2.99\\
MRN8 & MRN77 & 4977.81 & 4756 & 1.05 & 2.99\\
MRN$_{\rm fixed}$ & MRN77 & 7730.37 & 4885 & 1.58 & 3.19\\
MW, $R_{\rm V}=$3.1, Case A, $b_{\rm c}=$0 & WD01 & 5301.62  & 4772 & 1.11 & 3.45\\
MW, $R_{\rm V}=$3.1, Case A, $b_{\rm c}=$1 & WD01 & 5316.07  & 4772 & 1.11 & 3.46\\
MW, $R_{\rm V}=$3.1, Case A, $b_{\rm c}=$2 & WD01 & 5337.26  & 4772 & 1.12 & 3.47\\
MW, $R_{\rm V}=$3.1, Case A, $b_{\rm c}=$3 & WD01 & 5373.16  & 4772 & 1.13 & 3.50\\
MW, $R_{\rm V}=$3.1, Case A, $b_{\rm c}=$4 & WD01 & 5429.55  & 4772 & 1.14 & 3.51\\
MW, $R_{\rm V}=$3.1, Case A, $b_{\rm c}=$5 & WD01 & 5483.86  & 4772 & 1.15 & 3.54\\
MW, $R_{\rm V}=$3.1, Case A, $b_{\rm c}=$6 & WD01 & 5530.73  & 4772 & 1.16 & 3.54\\
MW, $R_{\rm V}=$4.0, Case A, $b_{\rm c}=$0 & WD01 & 5594.42  & 4772 & 1.17 & 3.58\\
MW, $R_{\rm V}=$4.0, Case A, $b_{\rm c}=$1 & WD01 & 5620.35  & 4772 & 1.18 & 3.60\\
MW, $R_{\rm V}=$4.0, Case A, $b_{\rm c}=$2 & WD01 & 5661.76  & 4772 & 1.19 & 3.62\\
MW, $R_{\rm V}=$4.0, Case A, $b_{\rm c}=$3 & WD01 & 5704.37  & 4772 & 1.20 & 3.65\\
MW, $R_{\rm V}=$4.0, Case A, $b_{\rm c}=$4 & WD01 & 5755.28  & 4772 & 1.21 & 3.68\\
MW, $R_{\rm V}=$5.5, Case A, $b_{\rm c}=$0 & WD01 & 5837.79  & 4772 & 1.22 & 3.72\\
MW, $R_{\rm V}=$5.5, Case A, $b_{\rm c}=$1 & WD01 & 5867.40  & 4772 & 1.23 & 3.74\\
MW, $R_{\rm V}=$5.5, Case A, $b_{\rm c}=$2 & WD01 & 5926.88  & 4772 & 1.24 & 3.78\\
MW, $R_{\rm V}=$5.5, Case A, $b_{\rm c}=$3 & WD01 & 6009.88  & 4772 & 1.26 & 3.82\\
MW, $R_{\rm V}=$4.0, Case B, $b_{\rm c}=$0 & WD01 & 5651.93  & 4772 & 1.18 & 3.60\\
MW, $R_{\rm V}=$4.0, Case B, $b_{\rm c}=$1 & WD01 & 5668.03  & 4772 & 1.19 & 3.61\\
MW, $R_{\rm V}=$4.0, Case B, $b_{\rm c}=$2 & WD01 & 5683.67  & 4772 & 1.19 & 3.63\\
MW, $R_{\rm V}=$4.0, Case B, $b_{\rm c}=$3 & WD01 & 5705.05  & 4772 & 1.20 & 3.64\\
MW, $R_{\rm V}=$4.0, Case B, $b_{\rm c}=$4 & WD01 & 5748.57  & 4772 & 1.20 & 3.67\\
MW, $R_{\rm V}=$5.5, Case B, $b_{\rm c}=$0 & WD01 & 5899.59  & 4772 & 1.24 & 3.75\\
MW, $R_{\rm V}=$5.5, Case B, $b_{\rm c}=$1 & WD01 & 5923.16  & 4772 & 1.24 & 3.77\\
MW, $R_{\rm V}=$5.5, Case B, $b_{\rm c}=$2 & WD01 & 5962.51  & 4772 & 1.25 & 3.80\\
MW, $R_{\rm V}=$5.5, Case B, $b_{\rm c}=$3 & WD01 & 6061.75  & 4772 & 1.27 & 3.87\\
LMC avg, $R_{\rm V}=$2.6,$b_{\rm c}=$0 & WD01 & 5207.70 & 4772 & 1.09 & 3.32\\
LMC avg, $R_{\rm V}=$2.6,$b_{\rm c}=$1 & WD01 & 5307.05 & 4772 & 1.11 & 3.36\\
LMC avg, $R_{\rm V}=$2.6,$b_{\rm c}=$2 & WD01 & 5453.87 & 4772 & 1.14 & 3.41\\
LMC reg.~2, $R_{\rm V}=$2.6,$b_{\rm c}=$0 & WD01 & 5256.43 & 4772 & 1.10 & 3.36\\
LMC reg.~2, $R_{\rm V}=$2.6,$b_{\rm c}=$1 & WD01 & 5310.40 & 4772 & 1.11 & 3.39\\
LMC reg.~2, $R_{\rm V}=$2.6,$b_{\rm c}=$2 & WD01 & 5387.13 & 4772 & 1.13 & 3.41\\
SMC bar, $R_{\rm V}=$2.9,$b_{\rm c}=$0 & WD01 & 5170.81 & 4772 & 1.08 & 3.26\\
BARE-GR-S   & ZDA04 &     5170.78  & 4772 & {1.08} & 3.39\\
BARE-FR-FG & ZDA04 &   5172.15  & 4772 & {1.08} & 3.39\\
BARE-GR-B   & ZDA04 &     5121.25  & 4772 & {1.07} & 3.36\\
BARE-AC-S   & ZDA04 &     5505.55  & 4772 & {1.15} & 3.47\\
BARE-AC-FG & ZDA04 &   5458.44  & 4772 & {1.14} & 3.45\\
BARE-AC-B    & ZDA04 &     5324.88  & 4772 & {1.12} & 3.40\\
COMP-GR-S   & ZDA04 &    5513.39  & 4772 & {1.16} & 3.64\\
COMP-GR-FG & ZDA04 &  5539.46  & 4772 & {1.16} & 3.65\\
COMP-GR-B    & ZDA04 &    5460.75  & 4772 & {1.14} & 3.62\\
COMP-AC-S    & ZDA04 &    5768.22  & 4772 & {1.21} & 3.56\\
COMP-AC-FG & ZDA04 &  5734.70  & 4772 & {1.20} & 3.56\\
COMP-AC-B    & ZDA04 &    6691.16  & 4772 & {1.40} & 3.50\\
COMP-NC-S   & ZDA04 &    7421.49  & 4772 & {1.56} & 3.39\\
COMP-NC-FG & ZDA04 &  6890.40  & 4772 & {1.44} & 3.44\\
COMP-NC-B   & ZDA04 &    7023.66  & 4772 & {1.47} & 3.27\\
WSD                & WSD01 & 5488.28 & 4772 & 1.15 & 3.45\\
\enddata
\tablenotetext{a}{Cross section slopes $\alpha_{\rm 1.5\,keV}$ are
  derived from powelaw fits to the differential cross section in the
  $500'' < \theta_{\rm sc} < 4500''$ range at an energy of 1.5 keV.}
\end{deluxetable}

\end{document}